\documentclass[12pt]{article}
\usepackage{amsmath,amssymb}
\usepackage{cite}
\usepackage{graphicx}

\newcommand{\ii}{\mathrm{i}}

\begin{document}

\title{Two novel classes of solvable many-body problems of goldfish
type with constraints}
\author{ Francesco Calogero\thanks{\texttt{francesco.calogero@roma1.infn.it,
francesco.calogero@uniroma1.it}} \\
{\small Dipartimento di Fisica, Universit\`a di Roma ``La
Sapienza'', 00185 Rome, Italy}\\[-8pt]{\small Istituto Nazionale di
Fisica Nucleare, Sezione di Roma.}\\[16pt] \and David G\'omez-Ullate
\thanks{\texttt{david.gomez-ullate@fis.ucm.es}}\\
{\small Departamento de F\'isica Te\'orica II}\\[-8pt]
{\small Universidad Complutense}\\[-8pt]
{\small  28040 Madrid, Spain.}
 }

\maketitle
\begin{abstract}
Two novel classes of many-body models with nonlinear interactions
``of goldfish type'' are introduced.\ They are \textit{solvable}
provided the initial data satisfy a single constraint (in one
case; in the other, two constraints): i. e., for such initial data
the solution of their initial-value problem can be achieved via
\textit{algebraic} operations, such as finding the eigenvalues of
given matrices or equivalently the zeros of known polynomials.
\textit{Entirely isochronous} versions of some of these models are
also exhibited: i.e., versions of these models whose nonsingular
solutions are \textit{all completely periodic} with the
\textit{same} period.
\end{abstract}
\newpage
\section{Introduction and main results}

Long time ago the possibility was noticed of using the \textit{nonlinear}
mapping among the zeros and the coefficients of a polynomial in order to
identify \textit{solvable} (classes of) many-body problems, characterized by
\textit{nonlinear }equations of motions of Newtonian type (with one-body and
two-body, generally velocity-dependent, forces) \cite{C1978,C2001}. [%
\textit{Terminology: }here and hereafter we denote as \textit{solvable} any
problem whose solution can be achieved by algebraic operations, such as
finding the zeros of polynomials]. The starting point of these developments
is a \textit{linear} PDE such as%
\begin{eqnarray}
\psi _{tt}-\left[ a_{1}-\left( N-1\right) \,a_{6} z\right] \,\psi
_{t}+\left[ a_{2}+a_{3}\,z-2\,\left( N-1\right)
\,a_{10}\,z^{\,2}\right] \,\psi _{z} &&
\notag \\
-\left( a_{4}+a_{5}\,z+a_{6}\,z^{\,2}\right) \,\psi _{zt}+\left(
a_{7}+a_{8}\,z+a_{9}\,z^{\,2}+a_{10}\,z^{\,3}\right) \,\psi _{zz} &&  \notag
\\
-N\,\left[ a_{3}+\left( N-1\right) \,\left( a_{9}-a_{10}\,z\right) \right]
\,\psi =0~. &&  \label{UrLinPDE}
\end{eqnarray}%
[\textit{Notation}: hereafter $N$ is an arbitrary positive integer ($N\geq 2$%
), the symbols $a_{j}$ and (see below) $A_{j},B_{j}$ denote (\textit{a priori%
} arbitrary) \textit{constants}, and subscripted variables denote
partial differentiations of the dependent variable $\psi( z,t) $;
note incidentally that this PDE coincides, up to trivial
notational changes, with eq. (2.3.3-1) of Ref. \cite{C2001}]. One
then notes that this PDE admits solutions which are (for all time)
a monic polynomial of degree $N$ in $z$, and introduces the $N$
zeros $z_{n}(t) $ and the $N$
coefficients $c_{m}(t) $ of such a polynomial solution,%
\begin{equation}
\psi \left( z,t\right) =\prod\limits_{n=1}^{N}\left[ z-z_{n}(t) %
\right] =z^{\,N}+\sum_{m=1}^{N}c_{m}(t) \,z^{\,N-m}~. \label{Map}
\end{equation}%
It is then easily seen that the fact that $\psi \left( z,t\right) $
satisfies the linear PDE\ (\ref{UrLinPDE}) implies that the coefficients $%
c_{m}( t) $ evolve according to the following system of \textit{%
linear} ODEs \cite{C2001}:
\begin{eqnarray}
&&\ddot{c}_{m}-\left( N+1-m\right) \,a_{4}\,\dot{c}_{m-1}-\left[
\left( N-m\right) \,a_{5}+a_{1}\right]
\,\dot{c}_{m}+m\,a_{6}\,\dot{c}_{m+1}
\notag \\
&&+\left( N+1-m\right) \,\left( N+2-m\right) \,a_{7}\,c_{m-2}+(
N+1-m) \,[ \left( N-m\right) \,a_{8}+a_{2}] \,c_{m-1}
\notag \\
&&-m\,\left[ \left( 2\,N-1-m\right) \,a_{9}+a_{3}\right]
\,c_{m}+m\,\left( m+1\right) \,a_{10}\,c_{m+1}=0~. \label{UrEqcm}
\end{eqnarray}%
Likewise, it can be shown (see Ref. \cite{C2001}, or Section 2
below) that the zeros $z_{n}(t) $ evolve according to the
following system
of \textit{nonlinear} PDEs%
\begin{eqnarray}
\ddot{z}_{n} &=&a_{1}\,\dot{z}_{n}+a_{2}+a_{3}\,z_{n}-2\,\left(
N-1\right)
\,a_{10}\,z_{n}^{\,2}  \notag \\
&&+\sum_{\substack{m=1\\m \ne n}}^N\left\{ \left(
z_{n}-z_{m}\right) ^{-1}\, \left[
2\,\dot{z}_{n}\,\dot{z}_{m}+\left( \dot{z}_{n}+\dot{z}_{m}\right)
\,\left( a_{4}+a_{5}\,z_{n}\right) \right. \right.  \notag \\
&&\left. \left. +a_{6}\,\left( \dot{z}_{n}\,z_{m}+\dot{z}_{m}\,z_{n}\right)
\,z_{n}+2\,\left(
a_{7}+a_{8}\,z_{n}+a_{9}\,z_{n}^{\,2}+a_{10}\,z_{n}^{\,3}\right) \right]
\right\} ~.  \label{UrEqzn}
\end{eqnarray}%
\textit{Notation }and \textit{terminology}: hereafter indices such as $m$, $%
n$ range from $1$ to $N$ (unless otherwise indicated), superimposed dots
denote differentiations with respect to the time $t$, and the $N$ zeros $%
z_{n}(t) $ are interpreted as the coordinates of $N$ ``point
particles'' evolving according to the ``Newtonian equations of motion'' (\ref%
{UrEqzn}) featuring one-body and two-body velocity dependent
forces; and an analogous interpretation shall be given in all
analogous cases below. Unless otherwise explicitly specified, we
hereafter assume that all quantities under consideration (except
the time $t$) are generally \textit{complex} numbers, so that the
motions of the particles characterized by the coordinates
$z_{n}(t) $ take place in the complex $z$-plane; of course such
motions can easily be reinterpreted as taking place in a ``more
physical'' \textit{real} plane, but we shall not devote space to
discuss this well-known aspect (see, for instance, Chapter 4 of
Ref. \cite{C2001}, entitled ``Solvable and or integrable many-body
problems in the plane, obtained by complexification'').

Clearly, since the parameters $a_{j}$ are (arbitrary)
\textit{constants}, the solution of the system of \textit{linear}
ODEs (\ref{UrEqcm}) satisfied by the coefficients $c_{m}(t) $ can
be achieved by purely \textit{algebraic} operations; and once the
$N$ coefficients $c_{m}(t) $ of the polynomial $\psi (z,t) $ have
been obtained, the computation of the $N$ zeros $z_{n}(t) $ of
this polynomial, see (\ref{Map}), is also a purely
\textit{algebraic} task. Hence the determination of the $N$
coordinates $z_{n}(t) $ whose time
evolution is characterized by the Newtonian equations of motion (\ref{UrEqzn}%
) can be achieved via purely \textit{algebraic} operations: this $N$-body
problem is \textit{solvable}.

The 10-parameter class of \textit{solvable} $N$-body models (\ref{UrEqzn})
is vast, and it includes several interesting cases \cite{C2001}. Even the
very simplest case, characterized by the vanishing of \textit{all} the
parameters $a_{j}$, is quite remarkable: in this case the Newtonian
equations of motion read simply
\begin{subequations}
\label{Gold}
\begin{equation}
\ddot{z}_{n}=\sum_{\substack{m=1\\m \ne n}}^{N}\frac{2\,\dot{z}_{n}\,\dot{z}_{m}%
}{z_{n}-z_{m}}~,  \label{Golda}
\end{equation}%
and the solution of the \textit{initial-value} problem for this
problem is given by the following simple proposition: \textit{the
}$N$\textit{\ coordinates }$z_{n}(t) $\textit{\ are the
}$N$\textit{\ zeros of
the following algebraic equation in }$z$,%
\begin{equation}
\sum_{n=1}^{N}\frac{\dot{z}_{n}\left( 0\right) }{z-z_{n}\left( 0\right) }=%
\frac{1}{t}~.  \label{Goldb}
\end{equation}%
Note that, by multiplying this equation by the product $\prod%
\limits_{n=1}^{N}\left[ z-z_{n}\left( 0\right) \right] ,$ it
becomes a polynomial equation, with time-dependent coefficients,
of degree $N$ in $z,$ which indeed generally has $N$ solutions, at
least if one allows the coordinates $z_{n}$ to be \textit{complex}
numbers, hence their time-evolution to take place in the
\textit{complex} $z$-plane; indeed such an evolution can also be
reinterpreted as the motion of $N$ point particles on a
\textit{real} (``physical'') plane, and its phenomenology is quite
amusing, see Section 2.4.2 of Ref. \cite{C2001}, entitled ``The
simplest model: explicit solution (the game of musical chairs),
Hamiltonian structure''. Because of these remarkable features, as
well as the neat form of the Newtonian equations of motion
(\ref{Golda}) (which also happen to be Hamiltonian), this $N$-body
model was given the honorary title of ``goldfish''
\cite{C2001b}, and this attribute was then extended to several \textit{%
solvable} models featuring in the right-hand side of their
Newtonian equations of motion (``acceleration equal force'') terms
such as that
appearing in the right-hand side of (\ref{Golda}): note that these \textit{%
solvable }models ``of goldfish type'' include now the models
belonging to the 10-parameter class (\ref{UrEqzn}) and several
models outside of this class (for an updated review of such models
see \cite{C2007}).

In the present paper we exhibit two classes of \textit{solvable }$N$-body
models ``of goldfish type'' which are, to the best of our knowledge, \textit{%
new}, hence, in particular, \textit{not} included in the class (\ref{UrEqzn}%
). These novel models are however \textit{solvable} only provided
the particle coordinates $z_{n}(t) $ satisfy \textit{additional}
conditions, which must of course be compatible with the time
evolution, so
that it is sufficient that they be satisfied, in the context of the \textit{%
initial-value} problem, by the \textit{initial data}.\thinspace This
limitation could be considered as a drawback of these new models, or instead
as a feature that adds to their interest in the context of mathematical
physics. The hunch that such models should exist was suggested to us by
recent results \cite{GKM2005,GKM2004a,GKM2006,GKM5,GKMnew} concerning
certain \emph{exceptional polynomial subspaces}, as tersely reviewed in
Appendix A.

The first class of these models is characterized by the following
(6-parameter) Newtonian equations of motion:
\end{subequations}
\begin{eqnarray}
\ddot{z}_{n} &=&A_{1}\,\dot{z}_{n}+A_{2}+A_{3}\,z_{n}-\frac{2\,A_{4}}{z_{n}}
\notag \\
&&+\sum_{\substack{m=1\\m \ne n}}^{N}\left\{ \left(
z_{n}-z_{m}\right) ^{-1}\,
\left[ 2\,\dot{z}_{n}\,\dot{z}_{m}+A_{5}\,\left( \dot{z}_{n}+\dot{z}%
_{m}\right) \,z_{n}\right. \right.  \notag \\
&&\left. \left. +2\,\left( A_{4}-A_{2}\,z_{n}+A_{6}\,z_{n}^{\,2}\right)
\right] \right\} ~.  \label{1Eqzn}
\end{eqnarray}%
Here the \textit{novel} element -- not included in the class
(\ref{UrEqzn}) -- is that associated with the constant $A_{4}$. In
the following section we show that these equations of motion
entail that the
coefficients $c_{m}(t) $ -- related to the coordinates $%
z_{n}(t) $ via the mapping (\ref{Map}) -- evolve according to the
following system of \textit{linear} ODEs:
\begin{eqnarray}
&&\ddot{c}_{m}-\left[ \left( N-m\right) \,A_{5}+A_{1}\right]
\,\dot{c}_{m}+\left( N-1-m\right) \,\left( N+2-m\right)
\,A_{4}\,c_{m-2} \notag \\&&\quad\,\, -\left( N+1-m\right)
\,\left( N-1-m\right) \,A_{2}\,c_{m-1}  \notag \\&&\quad\,\,
-\,m\,\left[ \left( 2\,N-1-m\right) \,A_{6}+A_{3}\right]
\,c_{m}=0~, \label{1Eqcm}
\end{eqnarray}%
of course with $c_{0}=1$ and $c_{m}=0$ for $m>N$ and $m<0$ (consistently
with (\ref{Map}), and also with this system of ODEs). Note that the \textit{%
linear} character of this system of ODEs entails the
\textit{solvable} character of the Newtonian equations of motion
(\ref{1Eqzn}). But these findings are only valid if the
coordinates $z_{n}(t) $ satisfy
the constraint%
\begin{equation}
\sum\limits_{n=1}^{N}\frac{1}{z_{n}(t) }=0~, \label{1Constraintz}
\end{equation}%
and correspondingly (see below) the coefficients $c_{m}(t) $
satisfy the constraint%
\begin{equation}
c_{N-1}(t) =0~.  \label{1Constraintc}
\end{equation}%
The treatment of Section 2 implies that these constraints, (\ref%
{1Constraintz}) respectively (\ref{1Constraintc}), are indeed compatible
with the evolution equations (\ref{1Eqzn}) respectively (\ref{1Eqcm}); hence
they are automatically satisfied, in the context of the \textit{initial-value%
} problem, provided the initial data for the $N$-body problem (\ref{1Eqzn})
satisfy the following two conditions:%
\begin{equation}
\sum\limits_{n=1}^{N}\frac{1}{z_{n}( 0) }=0~,~~~\sum%
\limits_{n=1}^{N}\frac{\dot{z}_{n}(0) }{\left[ z_{n}(0) \right]
^{\,2}}=0~,  \label{1znInitial}
\end{equation}%
and correspondingly the initial data for the \textit{linear }system of ODEs (%
\ref{1Eqcm}) satisfy the two conditions%
\begin{equation}
c_{N-1}(0) =\dot{c}_{N-1}(0) =0~.  \label{1cmInitial}
\end{equation}

Before turning to the second class of new models, let us display an
equivalent avatar of the Newtonian equations of motion (\ref{1Eqzn}):%
\begin{eqnarray}
\ddot{z}_{n} &=&\left( A_{1}+\frac{N-2}{2}\,A_{5}\right) \,\dot{z}%
_{n}-\left( N-2\right) \,A_{2}+\left[ A_{3}+\left( N-2\right) \,A_{6}\right]
\,z_{n}-\frac{2\,A_{4}}{z_{n}}  \notag \\
&&\!\!\!\!\!\!\!\!\!\!+\frac{A_{5}}{2}\,\dot{Z}+A_{6}\,Z+\sum_{\substack{m=1\\m
\ne n}}^{N}\left\{
\left( z_{n}-z_{m}\right) ^{-1}\,\big[ 2\,\dot{z}_{n}\,\dot{z}_{m}+\frac{%
A_{5}}{2}\,\left( \dot{z}_{n}+\dot{z}_{m}\right) \,\left(
z_{n}+z_{m}\right) \right.    \notag \\&&
\!\!\!\!\!\!\!\!\!\!\left. +2\,A_{4}-A_{2}\,\left(
z_{n}+z_{m}\right) +A_{6}\,\left( z_{n}^{\,2}+z_{m}^{\,2}\right)
\big] \right\} ~, \label{1EqznZ}
\end{eqnarray}%
where%
\begin{equation}
Z(t) =\sum_{n=1}^{N}z_{n}(t) ~.  \label{Z}
\end{equation}%
[This second version, (\ref{1EqznZ}) with (\ref{Z}), has the merit to
immediately yield (by summing over $n$ from $1$ to $N$ -- using (\ref%
{1Constraintz}) and the vanishing of the double sum in the right-hand side
due to the antisymmetry of the summand under the exchange of the dummy
indices $m$ and $n$) the linear ODE%
\begin{equation*}
\ddot{Z}=\left[ A_{1}+\left( N-1\right) \,A_{5}\right] \,\dot{Z}-N\,\left(
N-2\right) \,A_{2}+\left[ A_{3}+2\,\left( N-1\right) \,A_{6}\right] \,Z~;
\end{equation*}%
which is consistent, via the relation%
\begin{equation}
c_{1}(t) =-Z(t)
\end{equation}%
implied by (\ref{Map}), with the equation satisfied by $c_{1}(t) $
that obtains by setting $m=1$ in (\ref{1Eqcm}). On the other hand
the first version, (\ref{1Eqzn}), has the merit to feature only
one-body and two-body forces].

\bigskip

The second class of new models is characterized by the following
\mbox{(7-parameter)} Newtonian equations of motion:%
\begin{eqnarray}
\ddot{z}_{n} &=&B_{1}\,\dot{z}_{n}-\left( N-1\right)
\,B_{2}\,z_{n}-2\,\left( N-1\right) \,B_{3}\,z_{n}^{\,2}+B_{4}\,\frac{\dot{z}%
_{n}}{z_{n}}  \notag \\
&&+\sum_{\substack{m=1\\m \ne n}}^{N}\left\{ \left(
z_{n}-z_{m}\right) ^{-1}\, \left[
2\,\dot{z}_{n}\,\dot{z}_{m}+\left( \dot{z}_{n}+\dot{z}_{m}\right)
\,\left( B_{4}+B_{5}\,z_{n}\right) \right. \right.   \notag \\
&&\left. \left. \quad +\,B_{6}\,\left(
\dot{z}_{n}\,z_{m}+\dot{z}_{m}\,z_{n}\right)
\,z_{n}+2\,\left( B_{7}\,z_{n}+B_{2}\,z_{n}^{\,2}+B_{3}\,z_{n}^{\,3}\right) %
\right] \right\} ~.  \label{2Eqzn}
\end{eqnarray}

Clearly the \textit{novel} element (in (\ref{2Eqzn}) relative to (\ref%
{UrEqzn})) is now that associated with the constant $B_{4}$. In
the following section we show that these equations of motion,
(\ref{2Eqzn}), entail that the coefficients $c_{m}(t) $ -- related
to the coordinates $z_{n}(t) $ via the mapping (\ref{Map}) --
evolve as follows:
\begin{eqnarray}
\ddot{c}_{m}-\left( N-m\right) \,B_{4}\,\dot{c}_{m-1}-\left[ \left(
N-m\right) \,B_{5}+B_{1}\right] \,\dot{c}_{m}+m\,B_{6}\,\dot{c}_{m+1} &&
\notag \\
+\left( N+1-m\right) \,\left( N-m\right) \,B_{7}\,c_{m-1} &&  \notag \\
-m\,\left( N-m\right) \,B_{2}\,c_{m}+m\,\left( m+1\right)
\,B_{3}\,c_{m+1}=0~, &&  \label{2Eqcm}
\end{eqnarray}%
entailing the \textit{solvable} character of the Newtonian equations of
motion (\ref{1Eqzn}). But this conclusion is valid only if the coordinates $%
z_{n}(t) $ satisfy the constraint%
\begin{equation}
\sum\limits_{n=1}^{N}\frac{\dot{z}_{n}(t) }{z_{n}(t) }=0~,
\label{2Constraintz}
\end{equation}%
and correspondingly (see below) the coefficients $c_{m}(t) $
satisfy the constraint%
\begin{equation}
\dot{c}_{N}(t) =0~.  \label{2Constraintc}
\end{equation}%
The treatment of Section 2 implies that these constraints are indeed
compatible with the evolution equations (\ref{2Eqzn}) and (\ref{2Eqcm});
hence they are automatically satisfied, in the context of the \textit{%
initial-value} problem, provided the initial data for the $N$-body problem (%
\ref{2Eqzn}) satisfy the following single condition:%
\begin{equation}
\sum\limits_{n=1}^{N}\frac{\dot{z}_{n}(0) }{z_{n}(0) }=0~,
\label{2znInitial}
\end{equation}%
and correspondingly the initial data for the \textit{linear }system of ODEs (%
\ref{2Eqcm}) satisfy the single condition%
\begin{equation}
\dot{c}_{N}(0) =0~.  \label{2cmInitial}
\end{equation}

The paper is organized as follows. In Section 2 the results
reported above are proven. In
Sections 3 and 4 we discuss tersely the behavior of these two \textit{%
solvable }$N$-body problems, and in Section 5 \textit{entirely isochronous}
versions of these novel models are identified: they are characterized by the
fact that \textit{all} their nonsingular solutions are \textit{completely
periodic} (namely, periodic in \textit{all} their dependent variables $%
z_{n}(t) $) with the \textit{same} basic period (or possibly with
a -- generally not too large \cite{GS2005} -- \textit{integer
multiple} of it). The last Section 6 outlines tersely further
developments, to be reported in future publications. The paper is
completed by two Appendices:
in the first one certain relevant results concerning \textit{%
exceptional polynomial subspaces} are tersely reviewed, in the second one
certain computations are confined whose treatment in the body of the paper
would disrupt the flow of the presentation.

\bigskip

\section{Proofs}

The procedure to arrive at the results reported above is by now
textbook material \cite{C2001,C2007}, hence our treatment here can
be terse;
although the new twist should be emphasized, implying that the \textit{%
solvable} character of the new models applies only provided their
time evolution is somewhat restricted -- as already indicated in
the preceding section. The starting point of our treatment is the
\textit{linear} PDE (but see the
\textit{Remark} below)%
\begin{eqnarray}
&& \psi _{tt}-\left[ a_{1}-\left( N-1\right)z \,a_{6}\right]
\,\psi _{t}+\left[ a_{2}+a_{3}\,z-2\,\left( N-1\right)
\,a_{10}\right] \,\psi _{z}  \notag
\\&& \quad\,\,\,
-\left( a_{4}+a_{5}\,z+a_{6}\,z^{\,2}\right) \,\psi _{zt}+\left(
a_{7}+a_{8}\,z+a_{9}\,z^{\,2}+a_{10}\,z^{\,3}\right) \,\psi _{zz}
\notag
\\&&
\quad\,\,\,-\,N\,\left[ a_{3}+\left( N-1\right) \,\left(
a_{9}-a_{10}\,z\right) \right] \,\psi   \notag \\&&\qquad=\,
\frac{a_{11}}{z}\,\left[ \psi _{t}-\frac{\dot{c}_{N}}{c_{N}}\right] +\frac{%
a_{12}}{z}\,\left[ \psi _{z}-\frac{c_{N-1}}{c_{N}}\right].
\label{UrPDE}
\end{eqnarray}%
[\textit{Notation: }the symbols $a_{j}$ denote again (\textit{a
priori} arbitrary) constants, subscripted variables denote partial
differentiations, and the dependent variable $\psi (z,t) $, as
well as the coefficients $c_{N}(t) $ and $c_{N-1}(t) $, are
characterized by (\ref{Map})]. \vskip 0.2cm
\noindent\textit{Remark}. The consistency of this evolution PDE,
(\ref{UrPDE}), with the fact that the dependent variable $\psi
(z,t) $ is the monic polynomial (\ref{Map}) of degree $N$ in $z$
shall be clear from what follows. Note however that this evolution
equation, (\ref{UrPDE}), is a \textit{linear PDE} iff the two
constants $a_{11}$ and $a_{12}$ vanish (in which case it coincides
with (\ref{UrLinPDE})); otherwise it is in fact a
\textit{nonlinear functional} equation, as implied by the formulas
(clearly
entailed by (\ref{Map}))%
\begin{equation}
c_{N}(t) =\psi \left( 0,t\right) ~,~~~c_{N-1}(t) =\psi _{z}\left(
0,t\right) ~.  \label{cNcN-1psi}
\end{equation}

Via (\ref{Map}), this evolution equation (\ref{UrPDE})
entails\footnote{The derivation of the first of these two systems
of ODEs is a rather trivial exercise; the derivation of the second
system is as well trivial, but readers who find this second task
exceedingly painstaking are advised to use the formulas given in
Section 2.3.2 of Ref. \cite{C2001}, or (even more conveniently) in
Appendix A of Ref. \cite{C2007}.} the following (systems of)
evolution ODEs for the $N$ coefficients $c_{m}(t) $ and for the
$N$ zeros $z_{n}(t) $:
\begin{eqnarray}
\ddot{c}_{m}-\left( N+1-m\right) \,a_{4}\,\dot{c}_{m-1}-\left[ \left(
N-m\right) \,a_{5}+a_{1}\right] \,\dot{c}_{m}+m\,a_{6}\,\dot{c}_{m+1} &&
\notag \\
+\left( N+1-m\right) \,\left( N+2-m\right) \,a_{7}\,c_{m-2}+\left(
N+1-m\right) \,\left[ \left( N-m\right) \,a_{8}+a_{2}\right]
\,c_{m-1} &&
\notag \\
-m\,\left[ \left( 2\,N-1-m\right) \,a_{9}+a_{3}\right] \,c_{m}+m\,\left(
m+1\right) \,a_{10}\,c_{m+1} &&  \notag \\
=a_{11}\,\left( \dot{c}_{m-1}-\frac{\dot{c}_{N}}{c_{N}}\,c_{m-1}\right)
+a_{12}\left[ \left( N+2-m\right) \,c_{m-2}-\frac{c_{N-1}}{c_{N}}\,c_{m-1}%
\right], &&  \label{EvEqcm}
\end{eqnarray}%
\begin{eqnarray}
\ddot{z}_{n} &=&a_{1}\,\dot{z}_{n}+a_{2}+a_{3}\,z_{n}-2\,\left( N-1\right)
\,a_{10}\,z_{n}^{\,2}+a_{11}\,\frac{\dot{z}_{n}}{z_{n}}-\frac{a_{12}}{z_{n}}
\notag \\
&&+\sum_{\substack{m=1\\m \ne n}}^{N}\left\{ \left(
z_{n}-z_{m}\right) ^{-1}\, \left[
2\,\dot{z}_{n}\,\dot{z}_{m}+\left( \dot{z}_{n}+\dot{z}_{m}\right)
\,\left( a_{4}+a_{5}\,z_{n}\right) \right. \right.  \notag \\
&&\left. \left. +\,a_{6}\,\left(
\dot{z}_{n}\,z_{m}+\dot{z}_{m}\,z_{n}\right) \,z_{n}+2\,\left(
a_{7}+a_{8}\,z_{n}+a_{9}\,z_{n}^{\,2}+a_{10}\,z_{n}^{\,3}\right)
\right] \right\} ~.  \label{EvEqzn}
\end{eqnarray}%
 The observation that the system (\ref{EvEqcm})
is consistent with $c_{0}=1$ and $c_{m}=0$ for $m>N$ and $m<0$
confirms the consistency with the evolution equation (\ref{UrPDE})
of the assumption that $\psi (z,t) $ be a monic polynomial of
degree $N$ in $z,$ see (\ref{Map}).

In order to guarantee that these Newtonian equations of motion, (\ref{EvEqzn}%
), be \textit{solvable}, we must make sure that the corresponding evolution
equations (\ref{EvEqcm}) be effectively \textit{linear}. A straightforward
way to achieve this goal is to set to zero the two parameters $a_{11}$ and $%
a_{12}$; but this simply reproduces the previously known (class of) \textit{%
solvable }$N$-body models (\ref{UrEqzn}). There are however two other
possibilities, which do yield two \textit{new} classes of integrable systems.

The first possibility is to set $a_{11}=0$ (but $a_{12}\neq 0$) and to then
require that the system of evolution equations (\ref{EvEqcm}) be consistent
with the constraint (\ref{1Constraintc}). It is indeed plain that the system
(\ref{EvEqcm}) becomes then \textit{linear}:%
\begin{eqnarray}
\ddot{c}_{m}-\left( N+1-m\right) \,a_{4}\,\dot{c}_{m-1}-\left[ \left(
N-m\right) \,a_{5}+a_{1}\right] \,\dot{c}_{m}+m\,a_{6}\,\dot{c}_{m+1} &&
\notag \\
+\left( N+1-m\right) \,\left( N+2-m\right) \,a_{7}\,c_{m-2}+\left(
N+1-m\right) \,\left[ \left( N-m\right) \,a_{8}+a_{2}\right]
\,c_{m-1} &&
\notag \\
-m\,\left[ \left( 2\,N-1-m\right) \,a_{9}+a_{3}\right] \,c_{m}+m\,\left(
m+1\right) \,a_{10}\,c_{m+1} &&  \notag \\
=a_{12}\,\left( N+2-m\right) \,c_{m-2}, &&
\end{eqnarray}%
and it is moreover easily seen (by setting $m=N-1)$ that this system is
consistent with the constraint (\ref{1Constraintc}), provided the parameters
$a_{j}$ satisfy the following restrictions:%
\begin{equation}
a_{4}=a_{6}=2a_{7}-a_{12}=a_{8}+a_{2}=a_{10}=a_{11}=0~,
\end{equation}%
entailing that the system (\ref{EvEqcm}) becomes
\begin{eqnarray}
\ddot{c}_{m}-\left[ \left( N-m\right) \,a_{5}+a_{1}\right] \,\dot{c}_{m} &&
\notag \\
+\left( N-1-m\right) \,\left( N+2-m\right) \,a_{7}\,c_{m-2}-\left(
N+1-m\right) \,\left( N-1-m\right) \,a_{2}\,c_{m-1} &&  \notag \\
-m\,\left[ \left( 2\,N-1-m\right) \,a_{9}+a_{3}\right] \,c_{m}=0~, &&
\end{eqnarray}%
and likewise the system (\ref{EvEqzn}) becomes%
\begin{eqnarray}
\ddot{z}_{n} &=&a_{1}\,\dot{z}_{n}+a_{2}+a_{3}\,z_{n}-\frac{2\,a_{7}}{z_{n}}
\notag \\
&&+\sum_{\substack{m=1\\m \ne n}}^{N}\left\{ \left(
z_{n}-z_{m}\right) ^{-1}\, \left[
2\,\dot{z}_{n}\,\dot{z}_{m}+\left( \dot{z}_{n}+\dot{z}_{m}\right)
\,a_{5}\,z_{n}\right. \right.  \notag \\
&&\left. \left. +\,2\,\left(
a_{7}-a_{2}\,z_{n}+a_{9}\,z_{n}^{\,2}\right) \right] \right\} ~.
\end{eqnarray}%
It is easily seen that these systems coincide with (\ref{1Eqcm}) and (\ref%
{1Eqzn}) after the following trivial relabeling of the parameters: $%
a_{j}=A_{j}$ for $j=1,2,3,5;$ $a_{7}=A_{4},$ $a_{9}=A_{6}.$ And it
is moreover plain that the constraint (\ref{1Constraintc})
corresponds to the constraint (\ref{1Constraintz}), since the
relation (\ref{Map}) among the zeros $z_{n}(t) $ and the
coefficients $c_{m}(t) $ of the polynomial $\psi $ clearly
entails\footnote{Actually to reach these conclusions one needs the additional condition $%
c_{N}(t)\neq 0,$ namely (see (\ref{cNa})) $z_{n}(t) \neq 0;$ but
the fact that such a condition must be imposed initially, and that
it shall subsequently hold for all regular solutions of the
Newtonian equations of motion (\ref{1Eqzn}), is implied by the
presence of the term proportional to $A_{4}$ in the right-hand
side of these equations of motion.}
\begin{subequations}
\begin{equation}
c_{N}(t) =\left( -\right) ^{\,N}\,\prod\limits_{n=1}^{N}z_{n}(t)
~,~~~c_{N-1}(t) =\left( -\right)
^{\,N-1}\,\sum\limits_{m=1}^{N}\,\prod_{\substack{n=1\\n \ne m}}^N
z_{n}(t) ~,  \label{cNa}
\end{equation}%
hence%
\begin{equation}
c_{N-1}(t) =-c_{N}(t) \,\sum\limits_{n=1}^{N}\frac{1%
}{z_{n}(t) }~.  \label{cNb}
\end{equation}%

The second possibility is to set $a_{12}=0$ (but $a_{11}\neq 0$) and to then
require that the system of evolution equations (\ref{EvEqcm}) be consistent
with the constraint (\ref{2Constraintc}). It is indeed plain that the system
(\ref{EvEqcm}) becomes then \textit{linear}:
\end{subequations}
\begin{eqnarray}
\ddot{c}_{m}-\left[ \left( N+1-m\right) \,a_{4}-a_{11}\right] \,\dot{c}%
_{m-1}-\left[ \left( N-m\right) \,a_{5}+a_{1}\right] \,\dot{c}_{m}+m\,a_{6}\,%
\dot{c}_{m+1} &&  \notag \\
+\left( N+1-m\right) \,\left( N+2-m\right) \,a_{7}\,c_{m-2}+\left(
N+1-m\right) \,\left[ \left( N-m\right) \,a_{8}+a_{2}\right]
\,c_{m-1} &&
\notag \\
-m\,\left[ \left( 2\,N-1-m\right) \,a_{9}+a_{3}\right]
\,c_{m}+m\,\left( m+1\right) \,a_{10}\,c_{m+1}=0~,  &&
\end{eqnarray}%
and it is moreover easily seen (by setting $m=N)$ that this system is
consistent with the constraint (\ref{2Constraintc}), provided the parameters
$a_{j}$ satisfy the following restrictions:%
\begin{equation}
a_{4}-a_{11}=a_{7}=a_{2}=\left( N-1\right) \,a_{9}+a_{3}=a_{12}=0~,
\end{equation}%
entailing that the system (\ref{EvEqcm}) becomes%
\begin{eqnarray}
\ddot{c}_{m}-\left( N-m\right) \,a_{4}\,\dot{c}_{m-1}-\left[ \left(
N-m\right) \,a_{5}+a_{1}\right] \,\dot{c}_{m}+m\,a_{6}\,\dot{c}_{m+1} &&
\notag \\
+\left( N+1-m\right) \,\left( N-m\right) \,a_{8}\,c_{m-1} &&  \notag \\
-m\,\left( N-m\right) \,a_{9}\,c_{m}+m\,\left( m+1\right)
\,a_{10}\,c_{m+1}=0~, &&
\end{eqnarray}%
and likewise the system (\ref{EvEqzn}) becomes%
\begin{eqnarray}
\ddot{z}_{n} &=&a_{1}\,\dot{z}_{n}-\left( N-1\right)
\,a_{9}\,z_{n}-2\,\left( N-1\right) \,a_{10}\,z_{n}^{\,2}+a_{4}\,\frac{\dot{z%
}_{n}}{z_{n}}  \notag \\
&&+\sum_{\substack{m=1\\m \ne n}}^{N}\left\{ \left(
z_{n}-z_{m}\right) ^{-1}\, \left[
2\,\dot{z}_{n}\,\dot{z}_{m}+\left( \dot{z}_{n}+\dot{z}_{m}\right)
\,\left( a_{4}+a_{5}\,z_{n}\right) \right. \right.  \notag \\
&&\left. \left. +a_{6}\,\left( \dot{z}_{n}\,z_{m}+\dot{z}_{m}\,z_{n}\right)
\,z_{n}+2\,\left( a_{8}\,z_{n}+a_{9}\,z_{n}^{\,2}+a_{10}\,z_{n}^{\,3}\right) %
\right] \right\} ~.
\end{eqnarray}%
It is easily seen that these systems coincide with (\ref{2Eqcm}) and (\ref%
{2Eqzn}) after the following trivial relabeling of the parameters: $%
a_{j}=B_{j}$ for $j=1,4,5,6;$
$a_{8}=B_{7},\,\,a_{9}=B_{2},\,a_{10}=B_{3}.$ And it is moreover
plain\footnote{Again with the additional condition $c_{N}(t)\neq
0$, which need not be highlighted since the presence of the term
proportional to $B_{4}$ in the right-hand side of the Newtonian
equations of motion (\ref{2Eqzn}) guarantees automatically its
validity for all regular solutions of this system of ODEs.} that
the constraint (\ref{2Constraintc}) corresponds to the constraint
(\ref{2Constraintz}), see (\ref{cNa}).

\bigskip

\section{Behaviors of the $N$-body models belonging to the first class}

In this section we discuss -- in somewhat more detail than done in the
introductory Section 1 -- the behavior of the solutions of $N$-body models
belonging to the first class, see (\ref{1Eqzn}).

The \textit{general} solution of the \textit{linear }system of ODEs (\ref%
{1Eqcm}) with (\ref{1Constraintc}) is given by the formula%
\begin{equation}
c_{m}(t) =\sum_{\ell =1}^{N}\left[ \gamma _{\ell }^{\left(
+\right) }\,v_{m}^{\left( \ell \right) \left( +\right) }\,\exp
\left( \lambda _{\ell }^{\left( +\right) }\,t\right) +\gamma
_{\ell }^{\left( -\right) }\,v_{m}^{\left( \ell \right) \left(
-\right) }\,\exp \left( \lambda _{\ell }^{\left( -\right)
}\,t\right) \right] ~.  \label{Solcm}
\end{equation}%
Here the $N$-vectors $\underline{v}^{\left( \ell \right) \left( \pm \right) }
$ are the $2\,N$ eigenvectors corresponding to the $2\,N$ eigenvalues $%
\lambda _{\ell }^{\left( \pm \right) }$ of the (generalized) eigenvalue
equation%
\begin{equation}
\left( \lambda ^{\,2}\,\underline{\mathbf{1}}+\lambda \,\underline{M}%
^{(1)}\,+\underline{M}^{(2)}\right) \,\underline{v}=0~,  \label{EiEq}
\end{equation}%
where the $N\times N$ matrices $\underline{M}^{\left( 1\right) }$ and $%
\underline{M}^{\left( 2\right) }$ are defined componentwise as follows (see (%
\ref{1Eqcm})):
\begin{subequations}
\label{M}
\begin{equation}
M_{m,m}^{\left( 1\right) }=-\left[ \left( N-m\right) \,A_{5}+A_{1}\right] ~,
\label{Ma}
\end{equation}%
\begin{equation}
M_{m,m}^{\left( 2\right) }=-m\,\left[ \left( 2\,N-m-1\right) \,A_{6}+A_{3}%
\right] ~,
\end{equation}%
\begin{equation}
M_{m,m-1}^{\left( 2\right) }=-\left( N+1-m\right) \,\left( N-1-m\right)
\,A_{2}~,
\end{equation}%
\begin{equation}
M_{m,m-2}^{\left( 2\right) }=\left( N-1-m\right) \,\left( N+2-m\right)
\,A_{4}~,
\end{equation}%
with all other elements vanishing. As for the $2\,N$ constants $\gamma
_{\ell }^{\left( \pm \right) },$ they are \textit{arbitrary} except for the
two requirements
\end{subequations}
\begin{subequations}
\begin{equation}
\sum_{\ell =1}^{N}\left[ \gamma _{\ell }^{\left( +\right) }\,v_{N-1}^{\left(
\ell \right) \left( +\right) }+\gamma _{\ell }^{\left( -\right)
}\,v_{N-1}^{\left( \ell \right) \left( -\right) }\right] =0~,
\end{equation}%
\begin{equation}
\sum_{\ell =1}^{N}\left[ \gamma _{\ell }^{\left( +\right) }\,v_{N-1}^{\left(
\ell \right) \left( +\right) }\,\lambda _{\ell }^{\left( +\right) }+\gamma
_{\ell }^{\left( -\right) }\,v_{N-1}^{\left( \ell \right) \left( -\right)
}\,\lambda _{\ell }^{\left( -\right) }\right] =0~,
\end{equation}%
which clearly correspond to the constraints (\ref{1cmInitial}) via (\ref%
{Solcm}). Of course, in the context of the \textit{initial-value}
problem, the $2\,N$ constants $\gamma _{\ell }^{\left( \pm \right)
}$ are determined by the $2\,N$ initial data $c_{m}(0) ,$
$\dot{c}_{m}(0) $ (including (\ref{1cmInitial})) via the system of $2\,N$ \textit{%
linear }equations
\end{subequations}
\begin{subequations}
\begin{equation}
c_{m}(0) =\sum_{\ell =1}^{N}\left[ \gamma _{\ell }^{\left(
+\right) }\,v_{m}^{\left( \ell \right) \left( +\right) }+\gamma
_{\ell }^{\left( -\right) }\,v_{m}^{\left( \ell \right) \left(
-\right) }\right] ~,
\end{equation}%
\begin{equation}
\dot{c}_{m}(0) =\sum_{\ell =1}^{N}\left[ \gamma _{\ell }^{\left(
+\right) }\,v_{m}^{\left( \ell \right) \left( +\right) }\,\lambda
_{\ell }^{\left( +\right) }+\gamma _{\ell }^{\left( -\right)
}\,v_{m}^{\left( \ell \right) \left( -\right) }\,\lambda _{\ell
}^{\left( -\right) }\right] ~.
\end{equation}%
And in the context of the \textit{initial-value} problem for the Newtonian $N
$-body model (\ref{1Eqzn}) the initial values $c_{m}(0) ,$ $\dot{%
c}_{m}(0) $ are determined by the initial values $z_{n}(0)
,\,\dot{z}_{n}(0) $ via the following relations implied by
(\ref{Map}):
\end{subequations}
\begin{equation}
\prod\limits_{n=1}^{N}\left[ z-z_{n}(0) \right]
=z^{\,N}+\sum_{m=1}^{N}c_{m}(0) \,z^{\,N-m}~,
\end{equation}%
\begin{equation}
-\sum_{n=1}^{N}\dot{z}_{n}(0) \prod_{\substack{m=1\\m \ne n}}^N%
\left[ z-z_{m}(0) \right] =\sum_{m=1}^{N}\dot{c}_{m}(0)
\,z^{\,N-m}~.
\end{equation}

\noindent\textit{Remark}: let us emphasize that, while in the context of the \textit{%
initial-value} problem for the $N$-body problem (\ref{1Eqzn}) the constants $%
\gamma _{\ell }^{\left( \pm \right) }$ depend on the initial data (as just
explained), the $2\,N$ eigenvalues $\lambda _{\ell }^{\left( \pm \right) }$
-- characterizing via (\ref{Solcm}) the time evolution of the coefficients $%
c_{m}(t) $ of the polynomial $\psi (z,t) $ whose $N$ zeros
$z_{n}(t) $ yield the coordinates of the $N$ moving particles --
do \textit{not} depend on the initial data, but only on the
parameters $A_{j}$ that specify the $N$-body model (\ref{1Eqzn})
under consideration.

We now notice that the $N\times N$ matrix $\underline{M}^{\left( 1\right) }$
is \textit{diagonal}, and the $N\times N$ matrix $\underline{M}^{\left(
2\right) }$ is \textit{triangular} ($M_{nm}^{(2)}=0$ if $n<m$). Hence the
eigenvalues $\lambda _{m}^{\left( \pm \right) }$ are just the roots of the $%
N $ (decoupled) second-order equations%
\begin{equation}
\lambda ^{\,2}-\left[ \left( N-m\right) \,A_{5}+A_{1}\right] \,\lambda -m\,%
\left[ \left( 2\,N-m-1\right) \,A_{6}+A_{3}\right] =0~,  \label{Eigen}
\end{equation}%
i. e.
\begin{subequations}
\label{landa}
\begin{equation}
\lambda _{m}^{\left( \pm \right) }=\frac{1}{2}\,\left\{ \left(
N-m\right) \,A_{5}+A_{1}\pm \Delta_m \right\} ~,  \label{landaa}
\end{equation}%
\begin{eqnarray}
\Delta_m ^{\,2} &=&\left[ \left( N-m\right) \,A_{5}+A_{1}\right] ^{\,2}+4\,m\,%
\left[ \left( 2\,N-m-1\right) \,A_{6}+A_{3}\right]  \notag \\
&=&\left( N\,A_{5}+A_{1}\right) ^{\,2}+\left( A_{5}^{\,2}-4\,A_{6}\right)
\,m^{\,2}  \notag \\
&&+2\,\left[ 2\,A_{3}+2\,\left( 2\,N-1\right) \,A_{6}-A_{5}\,\left(
A_{1}+N\,A_{5}\right) \right] \,m~.  \label{delta}
\end{eqnarray}

The behavior of the solutions of the $N$-body problem (\ref{1Eqzn}) with (%
\ref{1Constraintz}) is given by the time evolution of the $N$ zeros $%
z_{n}(t) $ of the polynomial $\psi (z,t) $, see (\ref%
{Map}), whose coefficients evolve exponentially in time as entailed by (\ref%
{Solcm}) with (\ref{landa}). The study of the time evolution of
the particle coordinates $z_{n}(t) $ is therefore reduced to the
study of the motion of the zeros of a polynomial whose
coefficients depend exponentially on time. In the generic case --
characterized by exponents not all of which are purely imaginary
-- the asymptotic behavior of these zeros in the remote past and
future -- characterizing the qualitative behavior of the $N$-body
model under consideration -- can therefore be easily evinced by
the treatment provided in Appendix G (entitled ``Asymptotic
behavior of the zeros
of a polynomial whose coefficients diverge exponentially'') of Ref. \cite%
{C2001}. The subclass of $N$-body models (\ref{1Eqzn}) characterized by
parameters $A_{j}$ satisfying the following restrictions,
\end{subequations}
\begin{subequations}\label{confined}
\begin{equation}
A_{1}=\ii\,\alpha ~,~~~A_{3}=\gamma ~,~~~A_{5}=\ii\,\beta
~,~~~A_{6}=\eta ~,
\end{equation}%
\begin{equation}
\left[ \left( N-m\right) \,\beta +\alpha \right] ^{\,2}-4\,m\,\left[ \left(
2\,N-m-1\right) \,\eta +\gamma \right] >0~,~~\,m=1,...,N~,
\end{equation}%
(where the $4$ constants $\alpha ,\beta ,\gamma ,\eta $ are \textit{all real}%
) -- conditions which are necessary and sufficient to guarantee that \textit{%
all} the $2\,N$ eigenvalues $\lambda _{m}^{\left( \pm \right) }$ are \textit{%
imaginary} numbers -- is instead clearly characterized by the
property that \textit{all} motions are \textit{confined}. The more
special subcase in which \textit{all} the $2\,N$ eigenvalues
$\lambda _{m}^{\left( \pm \right) }$\textit{\ }are \textit{
integer multiples} of a single quantity $\ii\,\omega $ with
$\omega $ a \textit{positive} constant, $\omega
>0 $, so that \textit{all} the coefficients $c_{m}(t) $ are
\textit{periodic} with the \textit{same} period $T=2\,\pi \,/\,\omega $, is
discussed in Section 5.

\bigskip

\section{Behaviors of the $N$-body models belonging to the second class}

In this section we discuss -- in somewhat more detail than done in the
introductory Section 1; but rather tersely, to avoid repetitions of
developments already elaborated in the preceding Section 3 -- the behavior
of the solutions of $N$-body models belonging to the second class, see (\ref%
{2Eqzn}).

The \textit{general} solution of the \textit{linear }system of ODEs (\ref%
{2Eqcm}) with (\ref{2Constraintc}) is given by a formula entirely analogous
to (\ref{Solcm}), and the subsequent developments are also analogous to
those reported in the preceding section, except that now the $N\times N$
matrices $\underline{M}^{\left( 1\right) }$ and $\underline{M}^{\left(
2\right) }$ are defined componentwise as follows (see (\ref{2Eqcm})):
\end{subequations}
\begin{subequations}
\label{2M}
\begin{equation}
M_{m,m+1}^{\left( 1\right) }=m\,B_{6}~,
\end{equation}%
\begin{equation}
M_{m,m}^{\left( 1\right) }=-\left[ \left( N-m\right) \,B_{5}+B_{1}\right] ~,
\end{equation}%
\begin{equation}
M_{m,m-1}^{\left( 1\right) }=-\left( N-m\right) \,B_{4}~,
\end{equation}%
\begin{equation}
M_{m,m+1}^{\left( 2\right) }=m\,\left( m+1\right) \,B_{3}~,
\end{equation}%
\begin{equation}
M_{m,m}^{\left( 2\right) }=-m\,\left( N-m\right) \,B_{2}~,
\end{equation}%
\begin{equation}
M_{m,m-1}^{\left( 2\right) }=\left( N+1-m\right) \,\left( N-m\right)
\,B_{7}~,
\end{equation}%
with all other elements vanishing. As for the $2\,N$ constants $\gamma
_{\ell }^{\left( \pm \right) },$ they are again \textit{arbitrary} except
now for the single requirement
\end{subequations}
\begin{equation}
\sum_{\ell =1}^{N}\left[ \gamma _{\ell }^{\left( +\right) }\,v_{N}^{\left(
\ell \right) \left( +\right) }\,\lambda _{\ell }^{\left( +\right) }+\gamma
_{\ell }^{\left( -\right) }\,v_{N}^{\left( \ell \right) \left( -\right)
}\,\lambda _{\ell }^{\left( -\right) }\right] =0~,
\end{equation}%
which clearly corresponds to the constraint (\ref{2cmInitial}) via (\ref%
{Solcm}).

Since the $N\times N$ matrices $\underline{M}^{\left( 1\right) }$ and $%
\underline{M}^{\left( 2\right) }$ are now neither diagonal nor triangular,
the computations of the eigenvalues $\lambda _{m}^{\left( \pm \right) }$
cannot be done now in explicit form (in the general case, with the
coefficients $B_{j}$ appearing in (\ref{2M}) unrestricted). Therefore a
discussion of the actual behavior of the solutions of this second \textit{%
solvable} model cannot be done here in this general case to the same
explicit extent as in the case of the first class of models, treated in the
preceding section. There is however a subcase of this second model for which
the treatment becomes closely analogous to that of the preceding section.
Indeed clearly if%
\begin{equation}
B_{3}=B_{6}=0~,  \label{BBzero}
\end{equation}%
the two $N\times N$ matrices $\underline{M}^{\left( 1\right) }$ and $%
\underline{M}^{\left( 2\right) }$ become both \textit{triangular}, hence the
corresponding eigenvalue equation becomes again a second-order algebraic
equation that can be easily solved\footnote{The  $N\times N$ matrices $%
\underline{M}^{\left( 1\right) }$ and $\underline{M}^{\left(
2\right) }$ become triangular also in the case $B_{4}=B_{7}=0,$
but whenever $B_{4}$ vanishes the novelty of the case treated in
this paper disappears, so we do
not pursue this case. }. Indeed it is easily seen that if the restriction (\ref%
{BBzero}) holds, this second-order equation determining the exponents $%
\lambda _{m}^{\left( \pm \right) }$ that characterize the behavior of the
\textit{second} class of models \textit{coincides} with (a subcase of) the
second-order equation (\ref{Eigen}) determining the exponents $\lambda
_{m}^{\left( \pm \right) }$ that characterize the behavior of the \textit{%
first} class of models, provided one sets
\begin{equation}
A_{1}=B_{1}~,~~~A_{5}=B_{5}~,~~~A_{6}=B_{2}~,~~~A_{3}+\left( N-1\right)
\,A_{6}=0~,  \label{AB}
\end{equation}%
hence the discussion given above for the first class of models becomes
applicable to the second class of models provided they are restricted by the
condition (\ref{BBzero}). And this includes of course also the more special,
\textit{entirely isochronous}, subcase, as discussed in Section 5.

\bigskip

\section{Isochronous versions of the solvable models}

A current definition of \textit{isochronous} systems (see, for instance,
\cite{C2007}) attributes this property to any dynamical system that
possesses at least one \textit{open} (hence \textit{fully-dimensional})
region in its phase space within which \textit{all} solutions are \textit{%
completely periodic} (i. e., periodic in \textit{all} dependent
variables) with the \textit{same} fixed period (of course, the
period being independent of the initial data, as long as they stay
within that \textit{isochronicity} region). The class of
\textit{isochronous} systems is vast \cite{C2007}, and it includes
a large zoo of systems interpretable as $N$-body problems
characterized by \textit{autonomous} equations of motion of
Newtonian type. A class of such \textit{isochronous} systems is
characterized by the Newtonian equations of motion (see Example
4.1.2-3 in Ref. \cite{C2007})
\begin{subequations}
\label{GenIso}
\begin{equation}
\underline{\ddot{z}}=-i\,\omega \,\underline{\dot{z}}+\sum_{k=1}^{K}%
\underline{F}^{(-k)}\left( \underline{z},\underline{\dot{z}}+i\,\omega \,%
\underline{z}\right) ~.
\end{equation}%
Here underlined variables indicate $N$-vectors, $\omega $ is a \textit{%
nonvanishing real} constant, $K$ is an arbitrary \textit{positive} integer,
and the $K$ ($N$-vector-valued) functions $\underline{F}^{(-k)}\left(
\underline{z},\underline{\tilde{z}}\right) $ are required to be \textit{%
analytic} (but not necessarily \textit{holomorphic}) in all their
$2\,N$ arguments and to satisfy the scaling property
\begin{equation}
\underline{F}^{(-k)}\left( c\,\underline{z},\underline{\tilde{z}}\right)
=c^{-k}\,\underline{F}^{(-k)}\left( \underline{z},\underline{\tilde{z}}%
\right) ~,~~~k=1,...,K~.
\end{equation}%
Indeed it has been shown \cite{C2007} that these dynamical systems, (\ref%
{GenIso}), are \textit{isochronous}, possessing an open (hence \textit{%
fully-dimensional}) region in their phase space in which \textit{all} their
solutions are \textit{completely periodic} with period
\end{subequations}
\begin{equation}
T=\frac{2\,\pi }{\omega }~.  \label{T}
\end{equation}

It is easily seen that the system (\ref{EvEqzn}) belongs to this class, (\ref%
{GenIso}), with $K=1$, provided

\begin{eqnarray}
&& a_{1}=-(2N-1)\ii\,\omega,\qquad a_{2}=[a_{11}-(N-1)a_4]\ii
\omega,\notag\\
&& a_3=2(N-1)\omega^2,\qquad \,\,\,\,\,a_5=2 \ii\omega,\\
&& a_{6}=0,\qquad \qquad\qquad \quad\,\, a_8=\ii\omega a_4 , \notag\\
&& a_9=-\omega^2,\qquad\qquad\qquad a_{10}=0,\notag
\end{eqnarray}
while the constants $a_{4}, a_{7}, a_{11}$ and $a_{12}$ remain
\textit{arbitrary}. It is clear that both systems, (\ref{1Eqzn})
respectively (\ref{2Eqzn}), fall within this class, provided the
constants $A_{j}$ respectively $B_{j}$ featured by these two
\textit{solvable} models satisfy appropriate restrictions (whose
explicit determination can be left as a simple exercise for the
diligent reader).

But here we like to use the more restrictive definition of
\textit{entire
isochronicity} \cite{C2007} , stating that a dynamical system is \textit{%
entirely isochronous} if \textit{all} its nonsingular solutions are \textit{%
completely periodic} with period $T$, or possibly with a (not arbitrarily
large; indeed, generally rather small \cite{GS2005}) \textit{integer}
multiple of $T$ -- implying that \textit{all} nonsingular solutions of an
\textit{entirely isochronous} system are in fact \textit{completely periodic}
with the \textit{same} period, which however need not be the \textit{%
primitive} period for all of them; or, equivalently, that the
property of \textit{isochronicity} holds in the \textit{entire}
phase space (with the possible exceptions of a lower-dimensional
set of initial data yielding {\em singular} solutions).

One can then assert (see Appendix B for a proof) that the $N$-body problem (%
\ref{1Eqzn}) is indeed \textit{entirely isochronous }provided the $6$
constants $A_{j}$ it features satisfy the following $4$ restrictions:
\begin{subequations}
\label{IsoCond}
\begin{eqnarray}
A_{1} &=&\left( k_{1}-\,N\,k_{2}\right) \,\ii\omega ~, \\
A_{3} &=&\frac{1}2\left( k_{2}+k_{3}\right) \,\left[ \left(
N-\frac{1}{2}\right)
\,\left( k_{2}-k_{3}\right) -k_{1}\right] \,\omega ^{\,2}~, \\
A_{5} &=&k_{2}\,\ii\omega ~, \\
A_{6} &=&\frac{k_{3}^{\,2}-k_{2}^{\,2}}{4}\,\omega ^{\,2}~,
\end{eqnarray}%
where $\omega $ is a \textit{positive} constant, $\omega >0$ and the $3$
numbers $k_{1},k_{2},k_{3}$ are \textit{integers}, unrestricted (i. e.,
positive, negative or vanishing) except for the requirement

\begin{equation}
k_{1}\neq -m\,k_{3}\text{ \thinspace \thinspace for \thinspace \thinspace }%
m=1,...,N~.  \label{landif}
\end{equation}%
\end{subequations}
Note that the two constants $A_{2}$ and $A_{4}$ remain completely
arbitrary. Also note that, if the two integers $k_{1}$ and $k_{2}$
vanish -- entailing that $A_{1}$ and $A_{5}$ vanish,
$A_{1}=A_{5}=0,$ and $A_{3}=\left( 1-2N\right)A_{6},$ the only
remaining restriction on $A_{6}$ being that it be a
\textit{positive} real number, $A_{6}>0$ -- then the equations of
motion (\ref{1Eqzn}) become \textit{real} (of course provided
\textit{real}
values are also assigned to the two \textit{a priori arbitrary} constants $%
A_{2}$ and $A_{4}$).

But there is a second, different class of \textit{entirely isochronous}
systems, that is obtained from (a subclass of) the \textit{solvable} system (%
\ref{1Eqzn}) with (\ref{1Constraintz}) via the standard ``trick''
procedure (see for instance Ref. \cite{C2007}), as we now explain.
To this end we take
as starting point the following special case of the model (\ref{1Eqzn}):%
\begin{equation}
\zeta _{n}^{\prime \prime }=-\frac{2\,A_{4}}{\zeta _{n}}+2\,\sum%
\limits_{m=1,\,m\neq n}^{N}\frac{\zeta _{n}^{\prime }\,\zeta _{m}^{\prime
}+A_{4}}{\zeta _{n}-\zeta _{m}}~,  \label{1Eqzita}
\end{equation}%
which corresponds to (\ref{1Eqzn}) with \textit{all} the constants
$A_{j}$ set to zero except $A_{4}$ (and moreover the purely
notational replacement -- convenient for what shall immediately
follow -- of the dependent variables $z_{n}(t) $ with the
dependent variables $\zeta _{n}\left( \tau \right) $, and of the
independent variable $t$ with the independent variable $\tau $,
appended primes denoting of course differentiations with respect
to this new independent variable $\tau $). We
then set%
\begin{equation}
z_{n}(t) =\exp \left( -\ii\,\omega \,t\right) \,\zeta _{n}\left(
\tau \right) ~,~~~\tau =\frac{\exp \left( \ii\,\omega \,t\right)
-1}{\ii\,\omega }~,  \label{Trick}
\end{equation}%
with $\omega $ a \textit{positive} constant, and we thereby obtain the new $N
$-body model characterized by the (\textit{autonomous}) Newtonian equations
of motion%
\begin{eqnarray}
\ddot{z}_{n}=-\ii\,\omega \,\dot{z}_{n}-\frac{2\,A_{4}}{z_{n}} &&  \notag \\
+2\,\sum_{\substack{m=1\\m \ne n}}^{N}\left\{ \left(
z_{n}-z_{m}\right) ^{-1}\, \left[ (\dot z_n+\ii\omega z_n)(\dot
z_m+\ii\omega z_m)+A_{4}\right] \right\} ~. &&
\end{eqnarray}%
Due to the way this model has been obtained it is clear that it describes an
\textit{entirely isochronous} $N$-body problem -- provided its initial data
are constrained by the conditions (\ref{1znInitial}), guaranteeing the
validity of (\ref{1Constraintz}) (both for this model and for the model (\ref%
{1Eqzita})). It is also easy to verify that this model is
\textit{not} a special case of the \textit{entirely isochronous}
$N$-body problem identified above, i. e. (\ref{1Eqzn}) with
(\ref{IsoCond}) (and of course with (\ref{1Constraintz})).

Let us turn now to a discussion of {\em entirely isochronous}
variants of the second class of solvable models treated in this
paper. The identification of such models cannot be explicitly
achieved in the general case along the same lines as done above --
in the initial part of this section -- for models belonging to the
first class of solvable systems, because in this second case one
cannot obtain explicitly the exponents $\lambda_m^{(\pm)}$, as
explained in the preceding section (see the paragraph after
\eqref{EiEq} unless the restriction (\ref{BBzero}) holds, in which
case the treatment given in the first part of this section for the
\textit{first }class of system can be
immediately extended to the \textit{second} class via the relations (\ref{AB}%
), which however entail (see the last of these equations
(\ref{AB})) that the restrictions (\ref{IsoCond}) must now be
complemented by the additional restriction
\begin{equation}
k_{2}+k_{3}=0~~~\text{or}~~\ 2\,k_{1}=N(k_2-k_3).
\end{equation}

It is moreover again possible, also for the models of the second class, to
proceed via the standard trick procedure. Since this approach is quite
analogous to that described immediately above for the \textit{first} class,
we deal with it quite tersely.

The starting point is now the \textit{solvable} system (\ref{2Eqzn}) with
all constants vanishing except $B_{4},$ which we now write as follows:%
\begin{equation}
\zeta _{n}^{^{\prime \prime }}=B_{4}\,\frac{\zeta _{n}^{\prime }}{\zeta _{n}}%
+\sum_{m=1,m\neq n}^{N}\frac{2\,\zeta _{n}^{\prime }\,\zeta _{m}^{\prime
}+B_{4}\,\left( \zeta _{n}^{\prime }+\zeta _{m}^{\prime }\right) }{\zeta
_{n}-\zeta _{m}}~.
\end{equation}%
It is now again appropriate to use the version (\ref{Trick}) of the trick,
obtaining thereby the $N$-body problem -- clearly \textit{entirely
isochronous}, because of the way it is obtained -- characterized by the
following equations of motion:%

\begin{eqnarray}
\ddot{z}_{n}=-i\omega \dot{z}_{n}+B_{4}\,\frac{\dot{z}_{n}+i\omega z_{n}}{z_{n}%
}\,\, &&  \notag \\
+\sum_{\substack{m=1\\m \ne n}}^{N}\frac{2\left(
\dot{z}_{n}+i\omega
z_{n}\right) \left( \dot{z}_{m}+i\omega z_{m}\right) +B_{4}\left[ \dot{z}%
_{n}+\dot{z}_{m}+i\omega \left( z_{n}+z_{m}\right) \right]
}{z_{n}-z_{m}}~. &&
\end{eqnarray}%

\section{Numerical results}

In this Section we perform a numerical integration of the
equations of motion \eqref{1Eqzn} of the first model in order to
illustrate the previous findings. The numerical integration has
been done with an embedded Runge-Kutta method of order 8(5,3) with
automatic step size control, as developed by Prince and Dormand
\cite{numerical}. The integration and the graphical output have
been performed with the software {\sc Dynamics
solver}\footnote{This program is available at
\texttt{http://tp.lc.ehu.es/jma/ds/ds.html} } developed by J.
Aguirregabiria.

In order to simplify our numerical treatment we have set in
\eqref{1Eqzn} $N=3$, and $A_1=A_2=A_5=0$. Hence, our numerical
study focuses  on the following system of three nonlinear complex
ODEs of second order:
\begin{eqnarray}
\ddot{z}_{n} =A_{3}\,z_{n}-\frac{2\,A_{4}}{z_{n}}
+\sum_{\substack{m=1\\m \ne n}}^{3} ( z_{n}-z_{m})^{-1}\, [
2\,\dot{z}_{n}\,\dot{z}_{m}  +2(A_{4}+A_{6}\,z_{n}^{\,2}) ].
\label{eq:3body}
\end{eqnarray}%

\subsubsection*{Generic (chaotic) behaviour} When $A_4= 0$ this
system is solvable and it is a particular case of the family
discussed in Chapter 2.3.3 of \cite{C2001}. When $A_4\neq 0$ and
no restrictions are imposed on the initial data, the system
\eqref{eq:3body} corresponds to the motion of the zeros of a
$3^{\text{rd}}$ degree polynomial whose coefficients evolve in
time according to the nonlinear system \eqref{EvEqcm}. The system
behaves in general in a chaotic manner as it can be evinced from
the numerical results in Figures {\bf 1e}, {\bf 1f} and {\bf 2}.
The initial data for that integration have been chosen to be
\begin{eqnarray}
&&z_1(0)=1.5-2\ii,\qquad \dot z_1(0)=2-\ii \notag\\
&&z_2(0)=-1+ 0 \ii,\quad\quad \dot z_2(0)=-1+\ii \label{data-nonlinear}\\
&&z_3(0)=1.2+0.4\ii,\quad \,\dot z_3(0)=2.32-0.76\ii \notag
\end{eqnarray}
which do not satisfy the constraints \eqref{1znInitial},
 while the values of the coefficients are
\begin{equation}\label{param_quasi}
 A_3=-5.1 \pi^2,\qquad A_4=15,\qquad A_6=\pi^2.\end{equation}

\subsubsection*{Quasi-periodic behaviour}

However, for the same values of these parameters,
\eqref{param_quasi}, initial data can be chosen within the
algebraic submanifold defined by \eqref{1znInitial}. One such
assignment is
\begin{eqnarray}
&&z_1(0)=2-2\ii,\qquad\quad\!\! \dot z_1(0)=2-\ii, \notag\\
&&z_2(0)=-1+ 0 \ii,\quad\quad \dot z_2(0)=-1+\ii,\label{data-linear}\\
&&z_3(0)=1.2+0.4\ii,\quad \,\dot z_3(0)=2.32-0.76\ii. \notag
\end{eqnarray}
As discussed in Section 2, now the system \eqref{eq:3body} is {\em
solvable}, and it corresponds to the motion of the zeros of a
monic $3^{\text{rd}}$ degree polynomial whose coefficients evolve
in time according to the following linear system:
\begin{subequations}
\begin{eqnarray}
\ddot c_1-(4A_6+A_3 )c_1=0,\\
c_2=0,\\
\ddot c_3-2A_4c_1-3(2A_6+A_3)c_3=0.
\end{eqnarray}
\end{subequations}
The solution of this system is trivial and from \eqref{confined}
and the discussion in Section 4, we see that it suffices that
\[4A_6+A_3<0,\qquad 2A_6+A_3<0\] for the orbits to be confined.
As it follows from \eqref{Solcm} with \eqref{landa}, the
coefficients of the polynomial are in general quasi-periodic
functions of time, hence also the behaviour of the system
\eqref{1znInitial} is quasi-periodic as it can be seen in Figures
{\bf 1c}, {\bf 1d} and {\bf 2}, where we have again assigned the
values \eqref{param_quasi} of the parameters, and the same initial
data \eqref{data-linear}.

\subsubsection*{Isochronous behaviour}

Following the discussion of Section 5 and Appendix B, the family
of {\em solvable} problems discussed above includes a subclass of
{\em entirely isochronous} systems. They occur when all the
frequencies of the system are commensurable. A look at
\eqref{IsoCond} reveals that these conditions are met if we choose
\begin{equation}
A_1=A_5=0,\qquad A_3=-\frac{5}4\ k_3^2\,\omega^2,\qquad
A_6=\frac{1}{4}\, k_3^2\,\omega^2.
\end{equation}
We have set $k_3=1$ and $\omega=2\pi$ so that the fundamental
period is $T=1$. The choice of parameters is then
\begin{equation}
 A_3=-5 \pi^2,\qquad A_4=15,\qquad A_6=\pi^2.
\end{equation}
and the initial data are the same as in the previous case
\eqref{data-linear}. The resulting motions are shown in Figures
{\bf 1a}, {\bf 1b} and {\bf 2}.


\begin{figure}[h]
\begin{center}
\begin{tabular}[h]{cc}
\begin{tabular}{c}
\includegraphics[width=.4\textwidth]{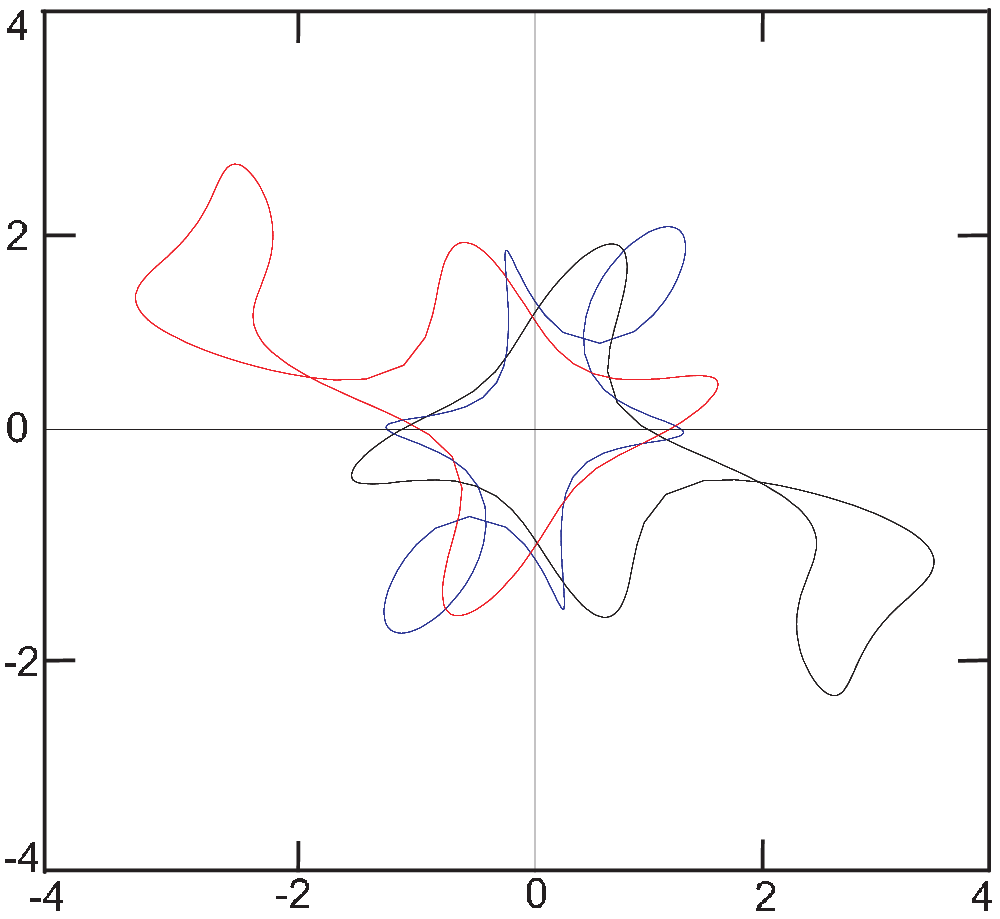}\\[-8pt] {\scriptsize{\bf Fig.1a }}
\end{tabular} &
\begin{tabular}{c}
\includegraphics[width=.4\textwidth]{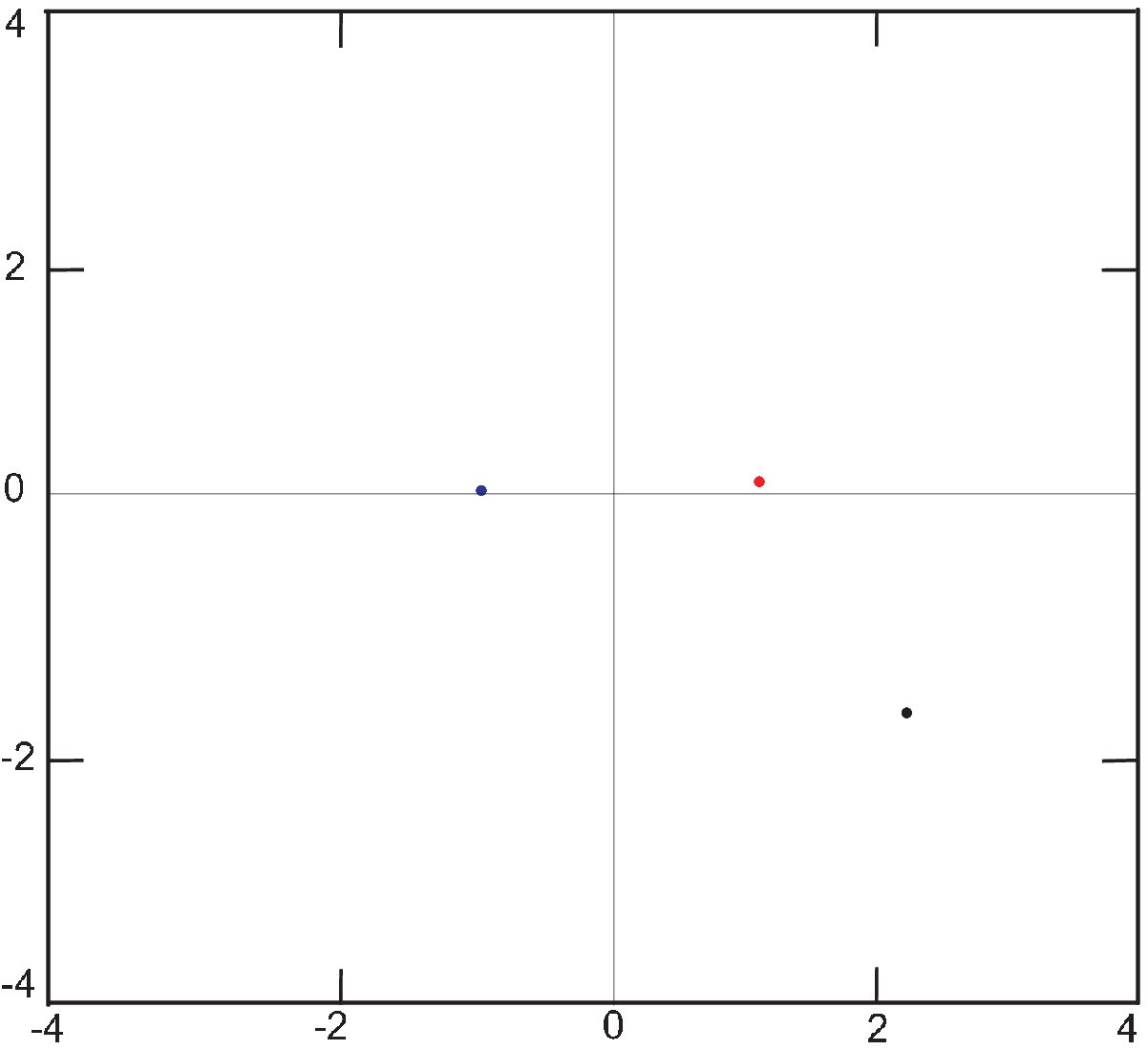}\\[-8pt] {\scriptsize{\bf Fig.1b } }
\end{tabular}\\[10pt]
\begin{tabular}{c}
\includegraphics[width=.4\textwidth]{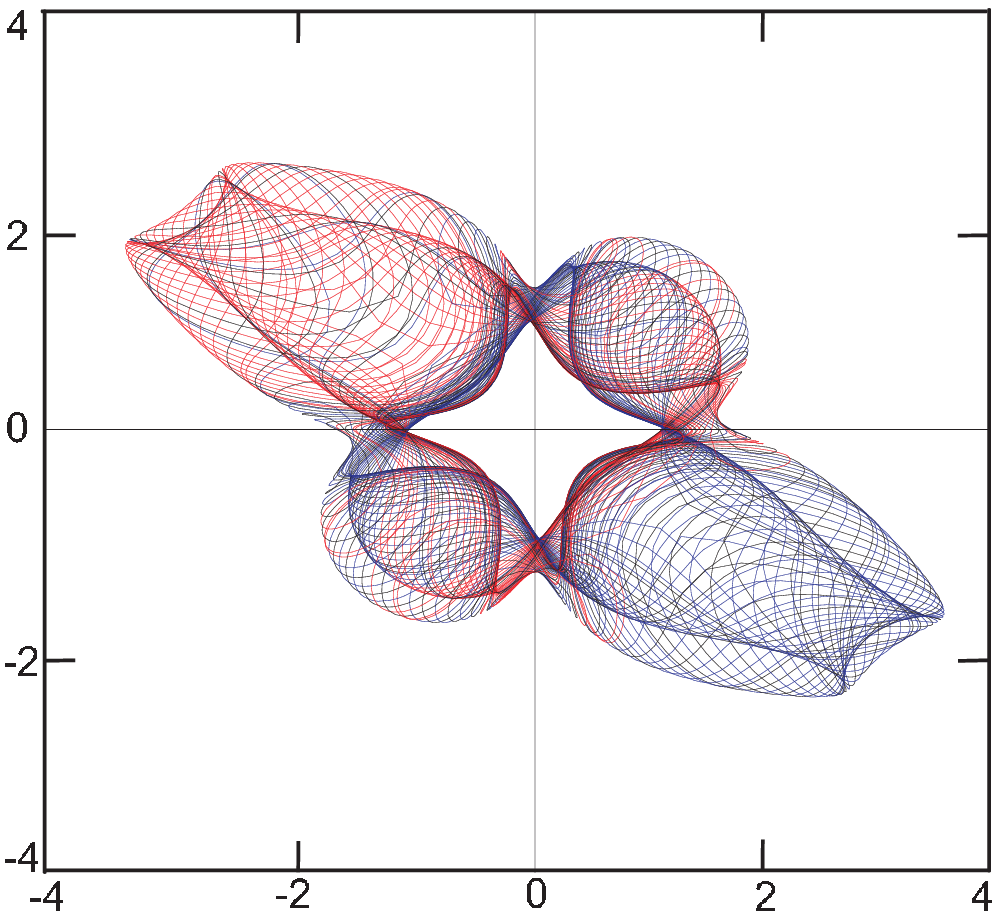}\\[-8pt] {\scriptsize{\bf Fig.1c } }
\end{tabular} &
\begin{tabular}{c}
\includegraphics[width=.4\textwidth]{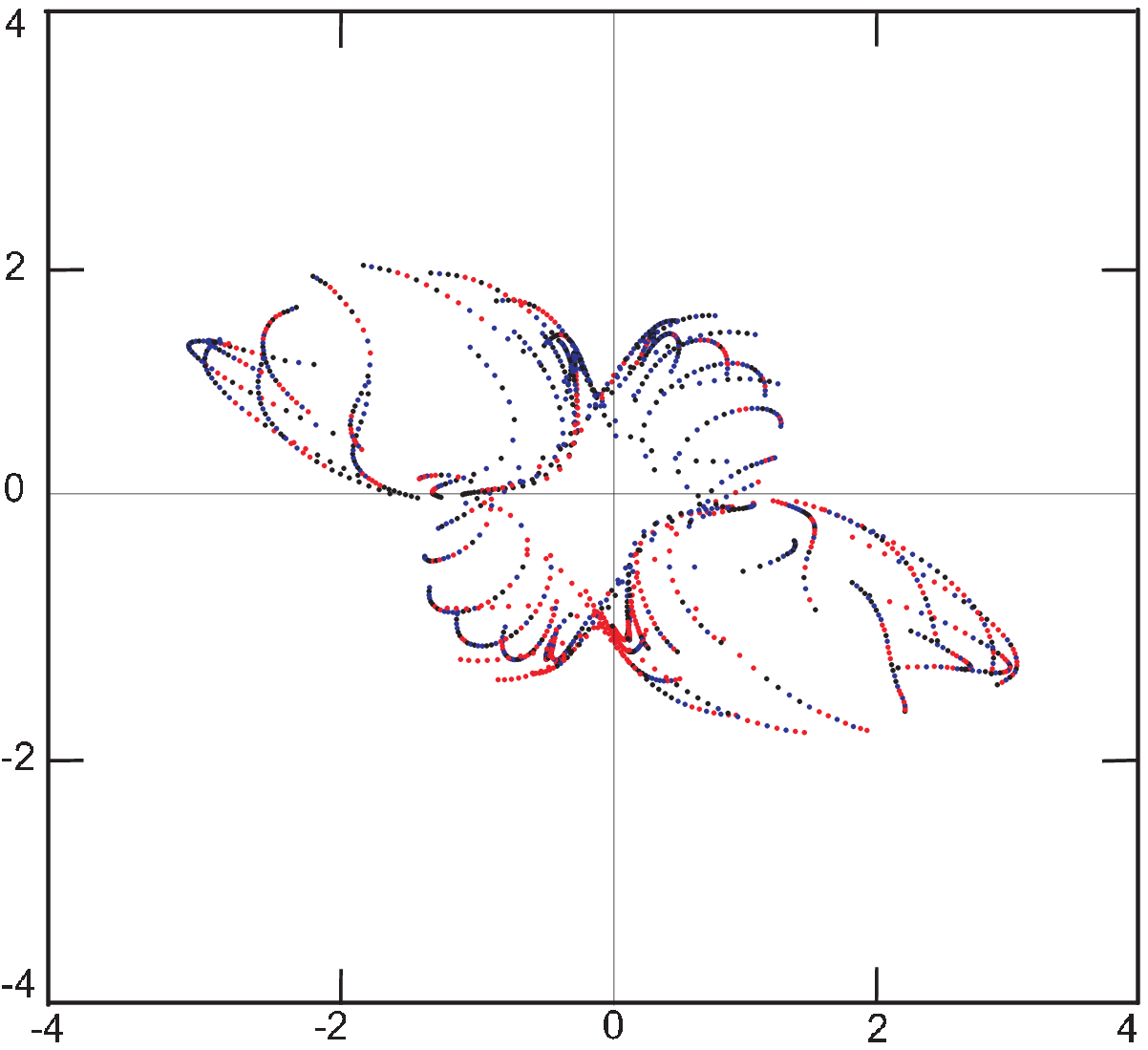}\\[-8pt] {\scriptsize{\bf Fig.1d } }
\end{tabular}\\[10pt]
\begin{tabular}{c}
\includegraphics[width=.4\textwidth]{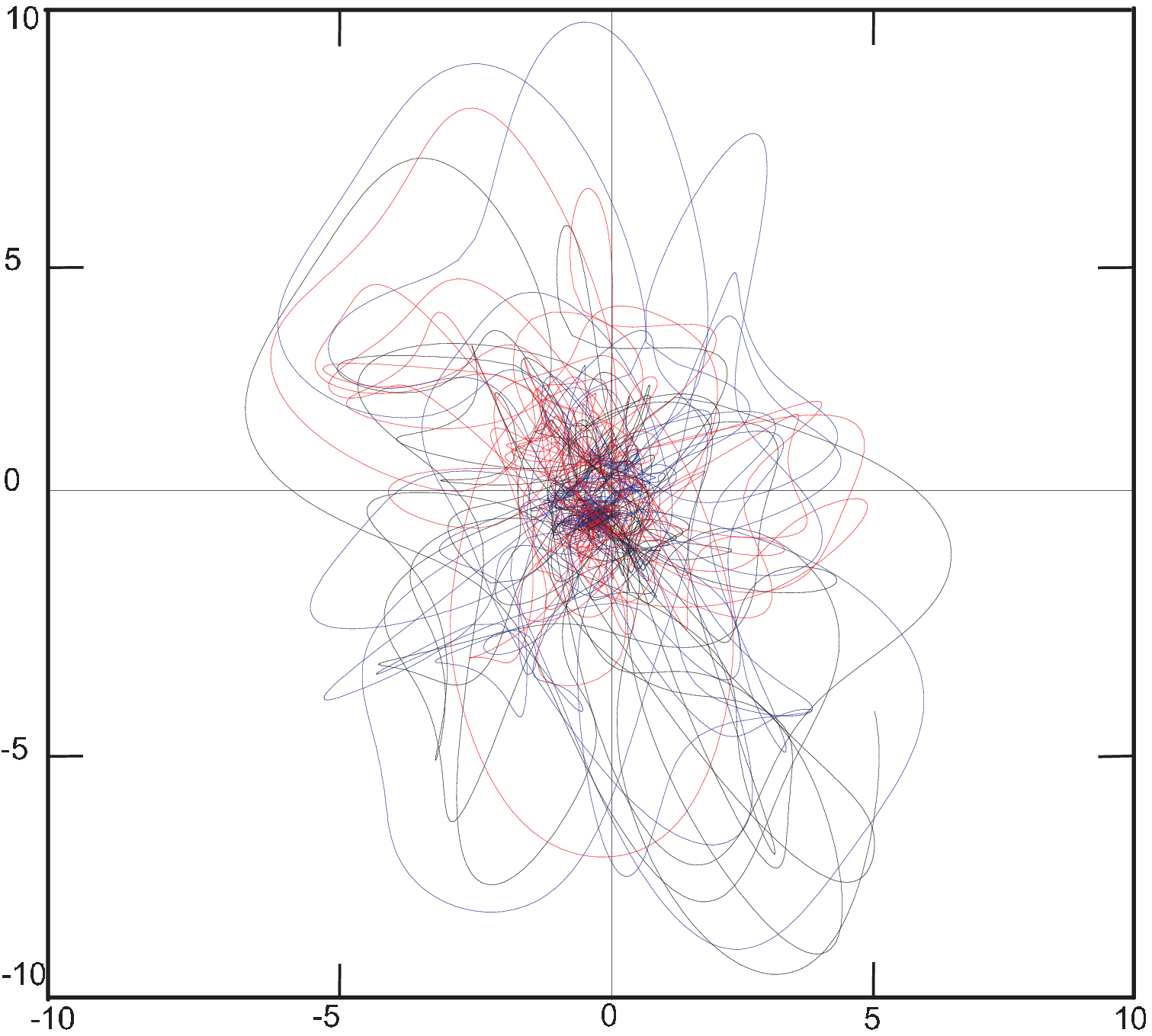}\\[-8pt] {\scriptsize{\bf Fig.1e } }
\end{tabular} &
\begin{tabular}{c}
\includegraphics[width=.4\textwidth]{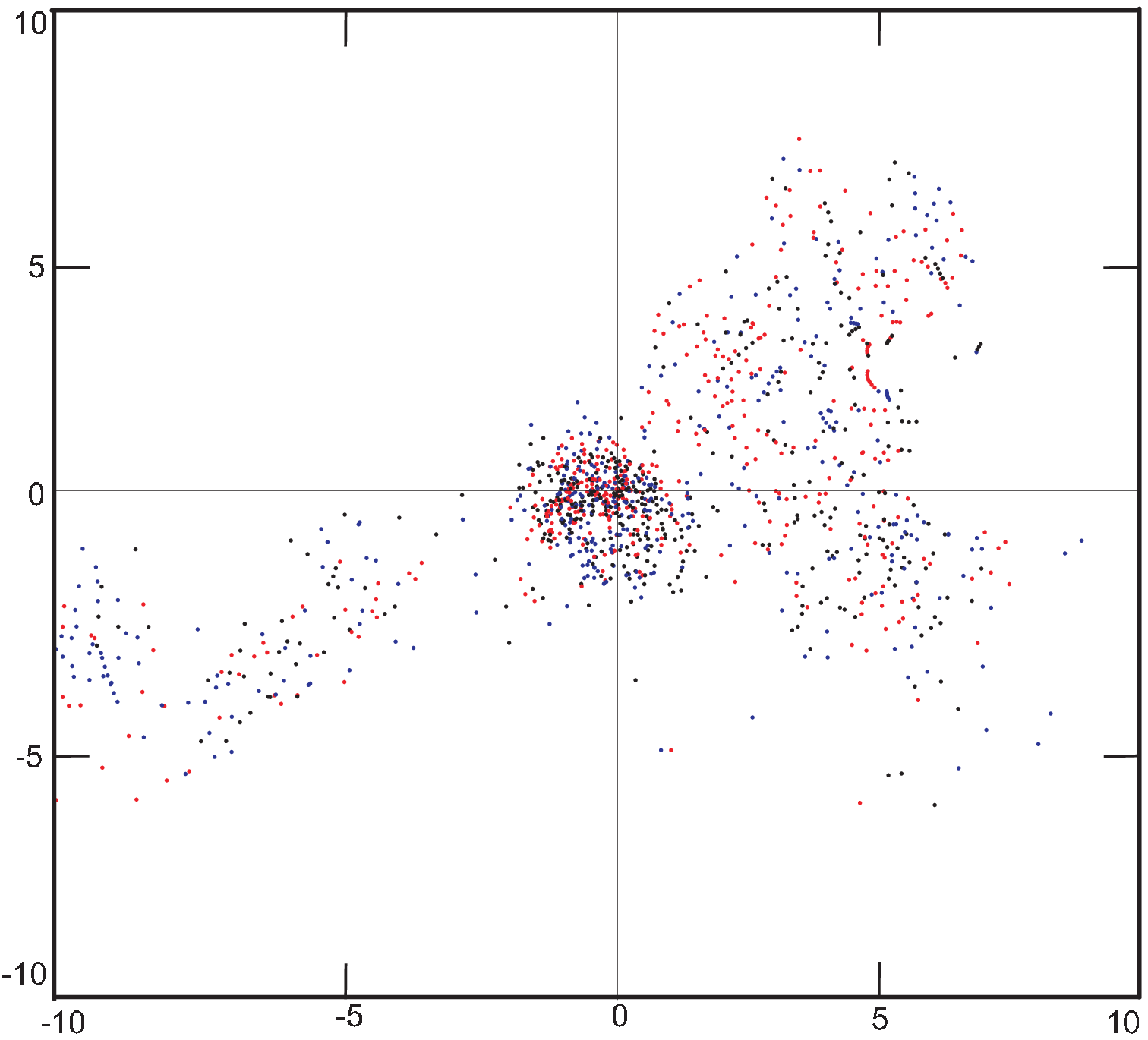}\\[-8pt] {\scriptsize{\bf Fig.1f } }
\end{tabular}
\end{tabular}
\end{center}
\caption{{\bf Left column:} trajectories of the $3$-body system
\eqref{eq:3body} in the periodic (1a), quasi-periodic (1c) or
chaotic regimes (1e). {\bf Right column:} Stereographic projection
of the trajectories: the positions of the three particles are
plotted at every integer multiple of $T=1$ up to $10^3$
iterations. }
\end{figure}

\begin{figure}[h]
\begin{center}
\includegraphics[width=.95\textwidth]{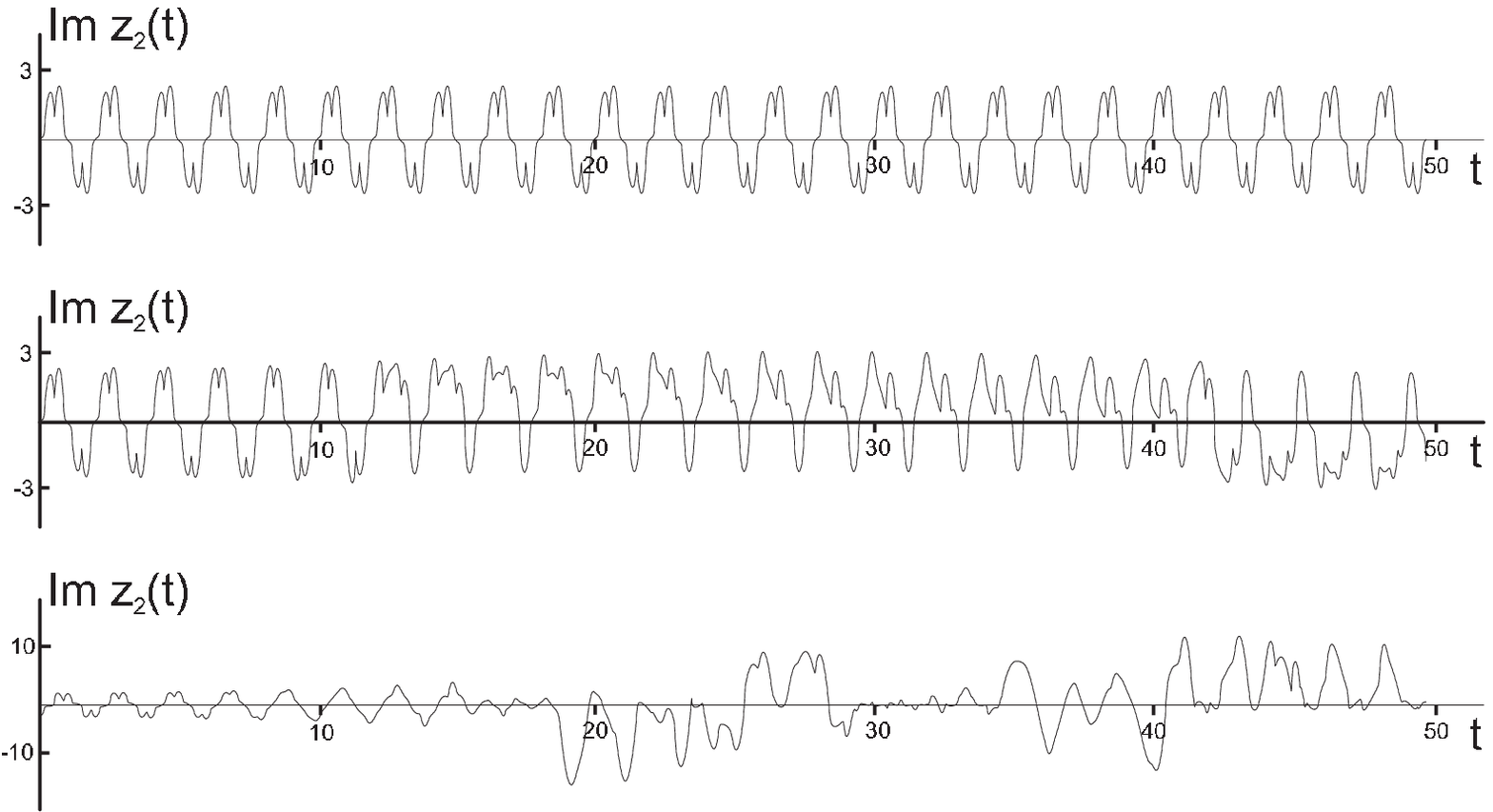}
\end{center}
\caption{A plot of $\text{Im}\,z_2(t)$ for the choice of
parameters and initial data described in Section 6. The periodic,
quasi-periodic and chaotic behaviours are manifest.}
\end{figure}

\section{Outlook}

The novel cases treated in this paper do not exhaust all the
possible {\em solvable} many-body problems associated to the
motion of the roots of polynomials whose coefficients satisfy a
linear system of ODEs. This question is related to the
classification of exceptional subspaces $X_{k}$ with co-dimension
$k>1$, which has not yet been completed. However, as mentioned in
Appendix A, some results based on $X_{2}$ spaces and operators
will be reported soon \cite{CG2006}.

\subsection*{Acknowledgements}

The results reported in this paper where obtained during a visit
in October 2006 of one of us (DGU) to the Department of Physics of
the University of Rome ``La Sapienza'', performed in the context
of the Collaboration Agreement among the University of Rome ``La
Sapienza'' and the Universidad Complutense of Madrid. The research
of DGU is supported in part by the Ram\'on y Cajal program of the
Ministerio de Ciencia y Tecnolog\'{i}a and by the DGI under grants
FIS2005-00752 and MTM2006-00478. It is a pleasure to acknowledge
illuminating discussions with Robert Milson.

\bigskip

\section*{Appendix A}

\bigskip

In this Appendix we explain the general method to write down
certain PDEs in 1+1 dimensions possessing solutions which are
monic polynomials $\psi (z,t) $ of degree $n$ in $z$ whose $N$
coefficients are time-dependent functions $c_{m}(t)$. The
requirement to ensure the solvability of the \textit{nonlinear}
$N$-body problem characterizing the time evolution of the $N$
zeros $z_{n}(t)$ of such a polynomial is that the coefficients
$c_{m}(t)$ evolve according to a \textit{linear} system of ODEs.

Let us denote by $\mathcal{P}_{N}$ the vector space of polynomials
in $z$ of degree less than or equal to $N$:
\begin{equation}
\mathcal P_N=\text{span}\left\{ 1,z,z^{2},\dots ,z^{N}\right\}
,\qquad \dim \mathcal P_N =N+1.
\end{equation}%
We shall denote also by $\mathcal{D}_{2}(\mathcal P_N)$ the vector
space of differential operators of order $2$ in $z$ with analytic
coefficients that leave the space $\mathcal P_N$ invariant, i.e.
\begin{equation}
\mathcal{D}_{2}(\mathcal P_N)=\{T=a_{2}(z)D_{zz}+a_{1}(z)D_{z}+a_{0}(z)\,|%
\,T\mathcal P_N\subset\mathcal P_N\}~.
\end{equation}%
\textit{Notation}: here and below we use the  notation $D_{z}\equiv \partial /\partial z,$ $%
D_{zz}\equiv \partial ^{2}/\partial z^{2},$ $D_{t}\equiv \partial
/\partial t,$ $D_{tt}\equiv \partial ^{2}/\partial t^{2},$
$D_{zt}\equiv \partial ^{2}/\partial z\partial t$.

Consider the action of the most general linear second-order
partial differential operator in $z$ and $t$:
\begin{equation}
L=a_{1}D_{tt}+a_{2}D_{zt}+a_{3}D_{zz}+a_{4}D_{z}+a_{5}D_{t}+a_{6},\qquad
a_i=a_i(z,t)\label{eq:T}
\end{equation}%
on the following time-dependent polynomial of degree $N$ in $z$ (see (\ref%
{Map})),%
\begin{equation}
\psi (z,t) =\prod\limits_{n=1}^{N}\left[ z-z_{n}(t) %
\right] =z^{N}+\sum_{m=1}^{N}c_{m}(t) \,z^{\,N-m}~.
\end{equation}

The restriction to second-order operators ensures that the {\em
nonlinear} many-body problem for the zeros will have at most
\textit{two-body} interactions, which is the more physically
interesting case. If many-body interactions are allowed, it is
obvious that the class of such solvable systems that can be
constructed with this method is larger. Since we want the
evolution of the coefficients $c_{m}(t)$ to be a \textit{linear
autonomous} (hence \textit{solvable}) system, we assume
$a_{i}(z,t)=a_{i}(z)$ and without loss of generality we also set
$a_{1}(z)=1$.

We now examine the conditions on $a_{i}(z)$ such that the equation $L[\psi
]=0$ implies a linear system for $c_{m}(t)$. These conditions arise from
imposing that $L$ acting on $\psi (z,t)$ produces a polynomial of degree $N-1
$. They are satisfied by a \textit{second-order} differential operator in $z$
and $t$ only in the following cases:

\begin{itemize}
\item the \textit{second-order} differential operator $D_{tt};$

\item one of the following 4 differential operators of
\textit{first-order in }$t$:
\begin{equation}
D_{t},~~~D_{zt},~~~zD_{zt},~~~z\left[ zD_{zt}-\left( N-1\right) D_{t}\right]
~;  \label{TimeDer}
\end{equation}

\item a \textit{second-order} differential operator in $z$ that
maps $\mathcal{P}_{N}$ to $\mathcal{P}_{N-1}.$
\end{itemize}

The characterization of vector spaces of linear (and nonlinear)
differential operators in one and several variables of any given
order that map $\mathcal{P}_{N}$ to $\mathcal{P}_{N-k}$ has been
treated in \cite{GKM2006}. Working out all the possible cases for
a second order differential operator in one variable that maps
$\mathcal{P}_{N}$ to $\mathcal{P}_{N-1}$ is a simple exercise that
produces the following result:
\begin{subequations}
\begin{eqnarray}
&&T_{2}^{+1}=z(zD_{z}-N+1)(zD_{z}-N+2), \\
&&T_{2}^{0}=z^{2}D_{zz}-N(N-1), \\
&&T_{2}^{-1}=zD_{zz}, \\
&&T_{2}^{-2}=D_{zz}, \\
&&T_{1}^{0}=zD_{z}-N, \\
&&T_{1}^{-1}=D_{z}~.
\end{eqnarray}
\end{subequations}
A linear combination of the eleven operators written above is
precisely equation \eqref{UrLinPDE} of this paper, which was
treated in Section 2.3.3 of \cite{C2001} (and several specific
cases were then investigated in Section 2.3.4). However, the
question arises whether this is the most general class of PDEs in
$z$ and $t$ which admit as solutions polynomials in $z$ whose
coefficients are functions of $t$ that evolve
according to a {\em linear} system. The recent discovery of the so-called \emph{%
exceptional polynomial subspaces of $\mathcal{P}_{N}$} shows that
other PDEs with those properties exist (and therefore other {\em
solvable} many-body problems) provided some constraints are
imposed.

An exceptional polynomial subspace $\mathcal{M}_{N}^{(k)}$ of
co-dimension $k$ in $\mathcal P_N$ is defined by the property that
some second-order differential operators that preserve
$\mathcal{M}_{N}^{(k)}$ \emph{do not
preserve} $\mathcal P_N$. More specifically, consider a space $\mathcal{M}%
_{N}^{(k)}\subset \mathcal P_N$ generated by $N+1-k$ linearly
independent
polynomials, all of them of degree at most $N$:%
\begin{equation}
\mathcal{M}_{N}^{(k)}=\text{span}\left\{ p_{1}(z),\dots
,p_{N+1-k}(z)\right\} ~.
\end{equation}%
We will say that $\mathcal{M}_{N}^{(k)}$ is an \textit{exceptional
polynomial
subspace} of co-dimension $k$ in $\mathcal P_N$ (for short, an $X_{k}$%
-space) if $\mathcal{D}_{2}(\mathcal M_{N}^{\left( k\right) })\not\subset \mathcal{D}_{2}(%
\mathcal P_N)$. Such exceptional subspaces do exist, and they
provide novel differential operators with polynomial
eigenfunctions. Exceptional subspaces have been first analyzed in
the context of quasi-exactly solvable potentials in quantum
mechanics \cite{GKM2005,GKM5}; they are also connected with the
Darboux transformation \cite{GKM2004a} and with non classical
families of orthogonal polynomials \cite{GKMnew}. $X_{1}$-spaces
have been fully classified arriving at the result that there is
essentially one such subspace up to projective transformations.
This space is precisely
\begin{subequations}
\begin{equation}
X_{1}=\text{span}\left\{ 1,z^{2},z^{3},\dots ,z^{N}\right\} ~,
\end{equation}%
which can also be characterized as
\begin{equation}
X_{1}=\{p\in \mathcal P_N\,|\,p^{\prime }(0)=0\}~.
\end{equation}%
It can be shown that $\dim \mathcal{D}_{2}(X_{1})=7$. The most
interesting element of this space is the operator
\end{subequations}
\begin{equation}
D_{zz}-\frac{2}{z}D_{z}~,  \label{eq:newop}
\end{equation}%
which preserves $X_{1}$ but not $\mathcal P_N$. The first class of
models treated in this paper -- see \eqref{1Eqzn}-- is related to
this exceptional space $X_{1}$, and the novel term proportional to
the constant $A_{4}$ in the many-body problem \eqref{1Eqzn} is
precisely that due to the inclusion
of this operator. The constraint on the many-body problem %
\eqref{1Constraintz} is precisely the one that defines the exceptional subspace $%
X_{1}$. The second class of many-body models treated above has a
different origin, not related to the exceptional subspaces of
$\mathcal P_N$, being instead associated with the role of the
time-derivative, see (\ref{TimeDer}).

Exceptional subspaces of higher co-dimension exist but a full
classification is not yet available. Some new many-body problems
with constraints associated to $X_{2}$-spaces will be treated in a
forthcoming publication \cite{CG2006}, their main novelty being
that the coefficients of the new differential operators are not
only inverse powers but rational functions of $z$.

\bigskip

\section*{Appendix B}

In this Appendix we justify the assertion made in Section 5, that
the conditions (\ref{IsoCond}) guarantee that the $N$-body model
(\ref{1Eqzn}) with (\ref{1znInitial}) is \textit{entirely
isochronous}.

Indeed, clearly this $N$-body problem is \textit{entirely isochronous} if
\textit{all }the solutions of the corresponding linear problem (\ref{1Eqcm})
are \textit{completely periodic} with the \textit{same} period $T$ (readers
for whom this is not clear are advised to consult, say, Ref. \cite{C2007}).
Clearly a necessary and sufficient condition for this to happen is that the $%
2\,N$ eigenvalues $\lambda _{m}^{\left( \pm \right) }$, see (\ref{landa}),
be \textit{all} \textit{integer multiples }of a common imaginary constant,
say
\begin{subequations}
\begin{equation}
\lambda _{m}^{\left( \pm \right) }=k_{m}^{\left( \pm \right)
}\,\ii\,\omega ~, \label{landak}
\end{equation}%
with $\omega $ a \textit{positive} constant, $\omega =2\,\pi \,/\,T>0$, and
the coefficients $k_{m}^{\left( \pm \right) }$ \textit{all integers}, and
moreover that%
\begin{equation}
\lambda _{m}^{\left( +\right) }\neq \lambda _{m}^{\left( -\right) }~,
\label{landal}
\end{equation}%
for all values of the index $m$, i. e. for $m=1,...,N$. Clearly the first of
these two relations, (\ref{landak}), can only be true if the quantity $%
\Delta_m ^{\,2}$, see (\ref{delta}), is a perfect square for all
values of $m,$ so that
\end{subequations}
\begin{equation}
\Delta_m =\alpha +\beta \,m~,  \label{Delta}
\end{equation}%
clearly entailing (see (\ref{delta}))
\begin{subequations}
\begin{equation}
\alpha =NA_{5}+A_{1}~,~~~\beta ^{\,2}=A_{5}^{\,2}-4A_{6}~,
\label{albe}
\end{equation}%
and (as we hereafter assume)%
\begin{equation}
2\,A_{3}+2\,\left( 2\,N-1\right) \,A_{6}-A_{5}\,\left( N\,A_{5}+A_{1}\right)
=\alpha \,\beta ~.  \label{A3}
\end{equation}%
Via (\ref{landa}) these formulas yield
\end{subequations}
\begin{equation}
\lambda _{m}^{\,\left( +\right) }=\alpha +\frac{m}{2}\,\left( \beta
-A_{5}\right) ~,~~~\lambda _{m}^{\,\left( -\right) }=-\frac{m}{2}\,\left(
\beta +A_{5}\right) ~.
\end{equation}%
Hence (\ref{landak}) entails%
\begin{equation}
\alpha =k_{1}\,\ii\,\omega ~,~~~\beta -A_{5}=2\,j\,\ii\,\omega
~,~~~\beta +A_{5}=2\,k\,\ii\,\omega ~,
\end{equation}%
with $k_{1},j,k$ \textit{integers}, and the (sum and difference of the) last
two formulas entail%
\begin{equation}
A_{5}=k_{2}\,\ii\,\omega ~,~~~\beta =k_{3}\,\ii\,\omega ~,
\end{equation}%
with $k_{2}$ and $k_{3}$ \textit{integers}. Hence the two equations (\ref%
{albe}) yield%
\begin{equation}
A_{1}=\left( k_{1}-\,N\,k_{2}\right) \,\ii\,\omega ~,~~~A_{6}=\frac{%
k_{3}^{\,2}-k_{2}^{\,2}}{4}\,\omega ^{\,2}~,
\end{equation}%
and from (\ref{A3}) one finally gets%
\begin{equation}
A_{3}=\frac{1}2\left( k_{2}+k_{3}\right) \,\left[ \left(
N-\frac{1}{2}\right) \,\left( k_{2}-k_{3}\right)
-k_{1}\right]\omega^2 ~.
\end{equation}%
The $4$ formulas (\ref{IsoCond}) are thereby proven. And it is moreover
clear that the condition (\ref{landal}) entails the requirement (\ref{landif}%
), thereby completing the proof of the results reported in Section 5.

\end{document}